\documentclass[11pt]{article}
\usepackage{epsfig}
\begin{document}

\def\wgta#1#2#3#4{\hbox{\rlap{\lower.35cm\hbox{$#1$}}
\hskip.2cm\rlap{\raise.25cm\hbox{$#2$}}
\rlap{\vrule width1.3cm height.4pt}
\hskip.55cm\rlap{\lower.6cm\hbox{\vrule width.4pt height1.2cm}}
\hskip.15cm
\rlap{\raise.25cm\hbox{$#3$}}\hskip.25cm\lower.35cm\hbox{$#4$}\hskip.6cm}}

\def\wgtb#1#2#3#4{\hbox{\rlap{\raise.25cm\hbox{$#2$}}
\hskip.2cm\rlap{\lower.35cm\hbox{$#1$}}
\rlap{\vrule width1.3cm height.4pt}
\hskip.55cm\rlap{\lower.6cm\hbox{\vrule width.4pt height1.2cm}}
\hskip.15cm
\rlap{\lower.35cm\hbox{$#4$}}\hskip.25cm\raise.25cm\hbox{$#3$}\hskip.6cm}}

\def\be{\begin{equation}}
\def\ee{\end{equation}}

\def\begeq{\begin{equation}}
\def\endeq{\end{equation}}
\def\zbar{\bar{z}}
\def\begeqar{\begin{eqnarray}}
\def\endeqar{\end{eqnarray}}
\def\partialbar{\bar{\partial}}
\def\wbar{\bar{w}}
\def\phibar{\bar{\phi}}
\def\xhat{\hat{x}}
\def\phat{\hat{p}}
\def\abar{\bar{a}}
\def\e{\epsilon}
\def\t{\theta}
\def\a{\alpha}
\def\k{\vec{k}}
\def\r{\vec{r}}

%
%
%
%

\title{The antiferromagnetic transition for 
the square-lattice Potts model}

\author{Jesper L.~Jacobsen$^{a,b}$ and Hubert Saleur$^{b,c}$\\
\smallskip\\
$^{a}$ Laboratoire de Physique Th\'eorique et Mod\`eles Statistiques \\
       Universit\'e Paris-Sud, B\^at.~100 \\ 
       91405 Orsay, France
\smallskip\\
$^{b}$ Service de Physique Th\'eorique\\
       CEA Saclay\\
       91191 Gif-sur-Yvette, France
\smallskip\\
$^{c}$ Department of Physics and Astronomy\\
University of Southern California\\
Los Angeles, CA 90089, USA
\smallskip\\
}

\maketitle

\begin{abstract}

  We solve in this paper the problem of the antiferromagnetic
  transition for the $Q$-state Potts model (defined geometrically for
  $Q$ generic using the loop/cluster expansion) on the square lattice.
  This solution is based on the detailed analysis of the Bethe ansatz
  equations (which involve staggered source terms of the type ``real''
  and ``anti-string'') and on extensive numerical diagonalization of
  transfer matrices. It involves subtle distinctions between the
  loop/cluster version of the model, and the associated RSOS and 
  (twisted) vertex models. The essential result is that the 
  twisted vertex model on the transition
  line has a continuum limit described by two bosons, one which is
  compact and twisted, and the other which is not, with a total central charge
  $c=2-{6\over t}$, for $\sqrt{Q}=2\cos{\pi\over t}$. The non-compact
  boson contributes a continuum component to the spectrum of critical
  exponents. For $Q$ generic, these properties are shared by the Potts
  model. For $Q$ a Beraha number, i.e., $Q=4\cos^{2}{\pi\over n}$ with
  $n$ integer, and in particular $Q$ integer, the continuum limit is
  given by a ``truncation'' of the two boson theory, and coincides
  essentially with the critical point of parafermions $Z_{n-2}$.

  Moreover, the vertex model, and, for $Q$ generic, the Potts model, exhibit a
  first-order critical point on the transition line---that is, the
  antiferromagnetic critical point is not only a point where
  correlations decay algebraically, but is also the locus of level
  crossings where  the derivatives of the free energy are discontinuous.
  In that sense, the thermal exponent of the Potts model is
  generically equal to $\nu={1\over 2}$. Things are however profoundly
  different for $Q$ a Beraha number. In this case, the
  antiferromagnetic transition is second order, with the thermal
  exponent determined by the dimension of the $\psi_{1}$ parafermion,
  $\nu={t-2\over 2}$. As one enters the adjacant ``Berker-Kadanoff''
  phase, the model flows, for $t$ odd, to a minimal model of CFT with
  central charge $c=1-{6\over (t-1)t}$, while for $t$ even it becomes
  massive. This provides a physical realization of a flow conjectured
  long ago by Fateev and Zamolodchikov in the context of $Z_{N}$
  integrable perturbations.

  Finally, though the bulk of the paper concentrates on the
  square-lattice model, we present arguments and numerical evidence
  that the antiferromagnetic transition occurs as well on other
  two-dimensional lattices.

\end{abstract}

\section{Introduction}

It is well known how to generalize the definition of the $Q$-state
Potts model to $Q$ real by turning to a loop/cluster formulation, and
the related six-vertex model (see, e.g., \cite{PaulMartin,SaleurI,PS}).
On the square lattice, the ferromagnetic Potts model thus defined
exhibits reasonably well understood critical properties. In many
respects, it can be considered as the ``lattice equivalent'' of the
Dotsenko-Fateev twisted bosonic theory, and thus, has served as a
benchmark for our understanding of numerous issues in conformal field
theory and integrable systems.

The antiferromagnetic case is much less understood. This is not too
surprising, as antiferromagnetic models often exhibit frustration, and
thus more complex phase diagrams. While the physical interest of this
model per se is probably limited to a few special values of
$Q$, it turns out that the difficulties encountered in its study are
very similar to those encountered in the study of models with
supergroup symmetries. These models are believed to play a crucial
role in the description of critical points in non-interacting
disordered systems.

Strong similarities with the $sl(2/1)$ spin chain relevant to the spin
quantum Hall effect \cite{EssFSal} in particular motivated us to
tackle the long overdue continuation of a first work on the topic by
one of us in 1991 \cite{SaleurI}. In this paper, we are finally able
to present a conjecture, supported by a long list of arguments and
partial calculations, for the conformal field theory describing the
critical antiferromagnetic Potts model on the square lattice for $Q\in
[0,4]$.  In a nutshell, this theory involves two bosons: a compact
boson not unlike the one appearing in the description of the
ferromagnetic critical point, {\sl plus} a non-compact boson. To our
knowledge, this is the second time only that a non-compact boson is
encountered in a statistical mechanics model with a finite number of
degrees of freedom per site, the first time being the super Goldstone
phases of \cite{JRS}.  The appearance of such bosons is generic in the
study of conformal field theories based on supergroups.

Another major peculiarity of the antiferromagnetic transition---and of
the neighboring ``Berker-Kadanoff'' phase---is the profound role
played by the quantum group symmmetry, and the related differences
between the Potts model and the vertex model for $Q$ generic, as well
as the difference between the case $Q$ generic and $Q$ a Beraha number
[i.e., $Q=4\cos^{2}{\pi\over n}$ with $n$ integer] for the Potts model.

Finally, the last peculiarity of the antiferromagnetic transition is 
that it is, for $Q$ generic, a first-order critical point, that is a 
point where correlations are algebraic but at the same time the 
derivatives of the free energy are discontinuous. This feature
was initially observed in Ref.~\cite{Sokal} for the particular
case of the $Q\to 0$ limit.

Obviously this quick summmary of our conclusions requires much
elaboration, and careful distinction between incarnations of the model
which are usually considered ``equivalent''. 
For easy reference, we review in section 2 key aspects of the various
  representations of the Potts model.  We also comment on the importance of
  boundary conditions  and on the
  definition of the transfer matrices in the various representations.
  Further background material  is given in the main text, and details
  can also be found in 
  \cite{SaleurI}.

The rest of the paper is organized as follows. We shall start in
section 3 by the simplest problem, which is the properties of the
(mostly untwisted) vertex model associated with the Potts model on
the antiferromagnetic critical line.

In section 4, we proceed to discuss the properties of the twisted
vertex model, as well as those of the loop model and thus the
$Q$-state Potts model for $Q$ generic.

In section 5, we discuss the related incarnation of RSOS models for 
$Q$ a Beraha number. We show how their continuum limit is given by a 
$Z_{k}$ parafermion theory, and use this in particular to complete 
the identification of the continuum limit of the antiferromagnetic  Potts 
model in section 6.

Another subtlety in this problem is the parity of the number of sites
(Potts spins) in the transfer matrix. While most of the paper
discusses the case $N$ even, interesting things also happen at $N$
odd, making the picture richer (the non-compact boson acquires a
twist). This is discussed in section 7.

Section 8 goes back to physics, with the discussion of the magnetic
exponent and the extremely delicate case of the thermal exponent. We
discuss there also the issue of the first-order critical point, and
the flow from parafermionic to minimal theories \`a la
Fateev-Zamolodchikov.

Section 9 contains conclusions as well as prospects for (much) future 
work. 

The appendix meanwhile discusses the limit $Q\rightarrow 0$ further,
in particular how the $R$ matrix for blocks of four vertices in the
staggered vertex model case reduces to the $osp(2/2)$ $R$ matrix in
this limit.

\section{Representations of the Potts model}

The  $Q$-state Potts model is defined initially
in terms of integer-valued spins $\sigma_i=1,2,\ldots,Q$ living at the
lattice vertices $i$, and interacting through nearest neighbour 
coupling,  with the  partition  function
\be
 Z = \sum_{\{\sigma_i\}} \prod_{\langle ij \rangle}
     {\rm e}^{K \delta(\sigma_i,\sigma_j)} \,.
\label{Zspin}
\ee
Here  $K$ is the coupling constant and $\langle ij \rangle$ denotes the
lattice edges. We shall refer to this as the {\em Potts spin representation}.

The low-temperature expansion of Eq.~(\ref{Zspin}) gives the {\em
cluster representation} \cite{Kasteleyn_69,Fortuin_72} for which
\be
 Z = \sum_{E \subseteq \langle ij \rangle}
     Q^{c(E)} ({\rm e}^K-1)^{|E|},
\label{Zcluster}
\ee
where $c(E)$ is the number of clusters (connected components) in the
edge subset $E$. The equivalence between the Potts spin representation and the 
cluster representation holds true with any boundary
conditions.  Note that now one can decide to 
give the variable  $Q$ any arbitrary real
value, hence extending the definition of the Potts model to $Q$
non-integer. 

Instead of the clusters, one may count the number of loops $l(E)$ of
the cluster boundaries which live on the medial, or surrounding,
lattice (here another square lattice oriented diagonally with respect
to the initial one). Using the Euler relation one obtains the {\em
loop representation} \cite{BKW_75}
\be
 Z = Q^{S/2} \sum_{E \subseteq \langle ij \rangle}
     Q^{l(E)/2+\eta(E)} \left( \frac{{\rm e}^K-1}{Q^{1/2}} \right)^{|E|}
\label{Zloop}
\ee
where $S$ is the number of lattice vertices. The function $\eta(E)=0$,
except on a torus if there is a cluster in $E$ that wraps around {\em
  both} periodic directions \cite{DFSZ}; in the latter case
$\eta(E)=1$. With this slight topological subtlety, one obtains a loop
representation of the Potts model for $Q$ arbitrary, totally
equivalent to the cluster representation. Formulas (\ref{Zcluster})
and (\ref{Zloop}) {\sl define} the model we want to study, and for
which we want to determine the phase boundaries, and critical
properties.

A key feature of the $Q$-state Potts model for $Q$ arbitrary is that
it has a non-local definition. While this is not a problem for
applications (e.g., the study of percolation, trees, forests, etc.),
it makes the analytical and numerical studies, or the construction of
associated conformal field theories, particularly tricky.  Let us see
this more explicitly.  We will be mostly interested in the case of the
square lattice, which we will often consider as being built up by a
{\sl transfer matrix} acting along the ``time'' ($t$) direction (of
length $M$ lattice spacings), the perpendicular direction being
referred to as the ``space'' ($x$) direction (of width $N$ lattice
spacings). We then have $S= N M$.

We consider for definiteness the case in which the transfer matrix
propagates along%
\footnote{In some cases it may be advantageous to study instead the
  case where the transfer matrix $T$ propagates diagonally with
  respect to the spin lattice (i.e., axially with respect to the
  medial lattice). \label{fnTM1}} the spin lattice (i.e., diagonally
with respect to the medial lattice), and we let the width of the
$x$-direction be $N$ Potts spins%
\footnote{In all subsequent figures of numerical data, the finite-size
  estimates of free energies $f$, central charges $c$, and physical
  scaling dimensions $x=h+\bar{h}$ are labelled by the width $N$ in
  terms of Potts spins. When the transfer direction is diagonal with
  respect to the spin lattice, we use the label $N^*$, meaning that
  the projected width is $\sqrt{2}N$ lattice spacings in terms of the
  spin lattice. \label{fnTM2}}.
 
The basis states for the transfer matrix in the Potts spin
representation, valid for $Q$ integer, are of course $N$ variables
$\sigma_i=1,2,\ldots,Q$, whence ${\rm dim}\ T=Q^N$. In the cluster
representation, basis states are set partitions of $N$ elements
respecting planarity, whence ${\rm dim}\ T=C_N$ with
$C_N=\frac{(2N)!}{N!(N+1)!}$ being Catalan numbers. For $N\gg 1$, we
have $C_N \sim 4^N$. Excited sectors are constructed by restricting to
set partitions with $\ge N_0$ components, of which exactly $N_0$ are
marked. By definition, the marked components cannot join among
themselves under the action by $T$.
  
In the loop representation, basis states are complete pairings of $2N$
points respecting planarity, with again ${\rm dim}\ T=C_N$. Excited
sectors are constructed by replacing $N_0$ of the pairings with
exactly $2N_0$ marked loop segments that cannot close among
themselves.
  

The crucial point is now that the partition functions (\ref{Zcluster})
or (\ref{Zloop}) cannot be written as traces of the transfer matrix
raised to the power $M$ because of the non-locality. This is of course
expected, the cluster or loop transfer matrices having different
dimension (and thus different sets of eigenvalues) than the Potts spin
transfer matrix for $Q$ integer. Instead, the partition functions are
written via more complicated ``functionals'' of the cluster or loop
transfer matrices, alllowing in particular for the disappearance of
many eigenvalues (corresponding to ``non-local'' information) for $Q$
integer.  For instance, with boundary conditions which are free in the
$x$-direction and periodic in the $t$-direction, the loop/cluster
partition function is given by $Z_M=\langle v_{\rm f} | T^M | v_{\rm
  i} \rangle$ for suitably initial and final vectors (and in fact with
a modified transfer matrix defined in terms of connectivities among
{\em two} time-slices) \cite{JS_05}. The end result is that
\be
 Z_M=\sum_{N_0=0}^N (2N_0+1)_q \sum_k 
 (\lambda^{(N_0)}_k)^M\label{Zexample},
\ee
where $\{\lambda^{(N_0)}_k\}$ is the set of eigenvalues of the 
foregoing transfer matrix, corresponding
to the sector with $N_0$ marked components defined above. The
amplitudes $(2N_0+1)_q$ are $q$-deformed numbers and are derived from
quantum group considerations \cite{PS}, while we have set 
$\sqrt{Q}=q+q^{-1}$. 

The essential point is that the amplitudes in (\ref{Zexample}) can
vanish when $Q$ is integer. In fact, setting $q=\exp(i\pi/t)$, they
can vanish whenever $t$ is rational. A remarkable feature of the Potts
model around the antiferromagnetic transition is that the disappearing
eigenvalues can be the leading ones, giving rise to strong
singularities in the thermodynamic properties. In contrast, in the
ferromagnetic region, the disappearing eigenvalues correspond only to
excited states, and lead merely to the disappearance of some
non-local observables from the spectrum of excitations.

To summarize, the lesson so far is that {\em for $q$ a root of unity,
  not all eigenvalues of the cluster or loop transfer matrix do
  contribute to the Potts partition function}. The same would of
course be true in the doubly free, or doubly periodic case.

It would now seem crucial to remove the non-local aspect of the
weights in the loop/cluster representation. To this end, we again
consider the case of free boundary conditions in the $x$-direction and
periodic in the $t$-direction. For all loops which are homotopic to a
point, assigning independently an orientation to each loop and giving
a weight $q^{\alpha/2\pi}$ to a loop that turns through an angle
$\alpha$ to the left then reproduces \cite{BKW_75} the full loop
weight of $Q^{1/2}$ upon setting $Q^{1/2}=q+q^{-1}$.  As for the
non-contractible loops, they would acquire instead a weight $2$ in this
formulation, but this can be corrected by assigning the weight $q$
(resp.\ $q^{-1}$) to any oriented loop that passes through the
periodic boundary condition.  Summing over the oriented loop
splittings of vertices which are compatible with given edge
orientations finally gives a representation in terms of the six-vertex
model \cite{BKW_75}. This model again lives on the medial lattice.
The Hamiltonian of the corresponding (XXZ) spin chain can be extracted
by taking the anisotropic limit, and is useful for studying the model
with the Bethe Ansatz technique. Note that the vertex model is in
general staggered, as the weight assigned to a given vertex depends on
whether this vertex stands on a direct or a dual edge of the original
spin lattice. Also, in the case of boundary conditions which are free
in the $x$-direction and periodic in the $t$-direction, the vertex model
has special weights for the vertices at the surface, corresponding to
boundary fields in the XXZ formulation. We shall refer to it as a
``twisted'' vertex model. An untwisted version would be obtained by
setting the boundary fields to zero.

Unlike for the cluster or loop transfer matrix, the transfer matrix of
the vertex model is a local object.%
\footnote{Note however that the presence of a seam in the twisted version
remains a non-local feature.}
Basis states are orientations of
$2N$ arrows, with an equal number of ups and downs, whence ${\rm dim}\
T=(N+1)C_N$. Excited sectors are constructed by having instead $N+k$
ups and $N-k$ downs. Periodic boundary conditions in the $x$-direction
are imposed by giving a special weight (twist) to the first arrow in
each row. But again, the correspondence with the Potts model implies
that the object of interest is not the trace of the transfer matrix
raised to the power $M$, but some functional of it. In the simplest
case of free boundary conditions in the $x$-direction and periodic in
the $t$-direction, adjusting the weight of non-contractible loops leads
to the expression $Z_M={\rm tr}\ q^{S^z} T^M$, where $S^z$ is the
conserved spin (arrow flux) along the transfer direction. The
commutation of the vertex transfer matrix with the quantum group
$SU(2)_{q}$ leads to a more compact expression of $Z_{M}$ in terms of
$q$-dimensions, with the same features of disappearing eigenvalues when
$q$ is a root of unity. So we get the same lesson that for $q$ a root
of unity, not all eigenvalues of the twisted vertex model transfer
matrix do contribute to the Potts partition function. The same would
of course be true in the doubly free, or doubly periodic case.

Obviously the phenomenon of disappearing eigenvalues requires great 
care in handling the results from transfer matrix analysis.

To complete this introduction we must now discuss briefly the doubly
periodic case. The correspondence between the models is then more
complicated.  Consider for instance the loop representation and its
vertex model reformulation.  The weight $Q^{1/2}$ for non-contractible
loops can only be reproduced, after orientation, by giving a special
weight that depends on the winding of the loop in both periodic
directions; this weight, in turn, transforms into a complex
combination of weighed sectors for the vertex model obtained after
summing over loop splittings! On top of this, the sector with no
non-contractible loops must also be singled out because of the extra
$Q^{\eta(E)}$ factor in (\ref{Zloop})! This is discussed in details in
the literature \cite{DFSZ}, and will be addressed below. For now, we
need to consider only what happens in the vertex model when the
non-contractible loops wind in only one direction.  We have already seen
what to do when they wind in the $t$-direction: their weight must be
adjusted from $2$ to $Q^{1/2}$ by introducing a modified trace term.
Similarly, when they wind in the $x$-direction, one must introduce a
twist term assigning the weight $q$ (resp. $q^{-1}$) to any oriented
loop passing through the $x$-periodic boundary condition.  This leads,
in the XXZ limit, to a modified coupling between the first and the
last $SU(2)_{q}$ spins, and defines what we call the {\em twisted
  vertex model or spin chain transfer matrix}. This is mostly an
intermediate object for us: while we consider the Potts partition
function or the untwisted vertex model partition function in this
paper, we do not enter the discussion of what would be the proper
definition of the twisted vertex model partition function.  Going
beyond the simple case of loops which are non-contractible in the
$x$-direction is a very technical topic \cite{ReadS} (best handled in the
framework of Temperley-Lieb and periodic Temperley-Lieb algebras)
which we will address when necessary in the following. Suffice it to
say that {\em for generic $q$, the eigenvalues of the cluster or loop
  transfer matrix are a subset of the eigenvalues of the vertex model
  transfer matrix for various twist sectors}.

\section{Bethe Ansatz analysis of the vertex model transfer matrix}

The antiferromagnetic critical line for the square-lattice Potts model
(defined, for $Q$ generic, using the loop/cluster formulation) was
identified by Baxter \cite{BaxterII} as
\begin{equation}
  \left({\rm e}^{K_{1}}+1\right)\left({\rm e}^{K_{2}}+1\right)=4-Q
  \label{antiferro}
\end{equation}
where $K_{1},K_{2}$ are the usual horizontal and vertical couplings. 
It is as always convenient to introduce the variables 
\begin{equation}
  x\equiv {e^{K}-1\over\sqrt{Q}}
\end{equation}
and set 
\begin{equation}
  x_{1}={\sin u\over\sin(\gamma-u)}
\end{equation}
where $\sqrt{Q}=2\cos\gamma$, with $\gamma\in \left[0,{\pi\over 
2}\right]$, and $u$ is a spectral parameter. 
The antiferromagnetic  line then corresponds to
\begin{equation}
  {1\over x_{2}}=-{\cos u\over\cos(\gamma-u)}\equiv{\sin 
    v\over\sin(\gamma-v)},~~v=u\pm{\pi\over 2}
\end{equation}
The Potts model on the line (\ref{antiferro}) 
is never selfdual. Selfduality  would  correspond instead  to 
$x_{1}={1\over x_{2}}$,  that is $u=v$.

The Potts model is generally related
to a six-vertex model with quantum group symmetry obtained from the
loop formulation by putting arrows on the medial lattice (see section
2, and Ref.~\cite{SaleurI} for further review), and using the spectral
parameters $u,v$ to define the vertex $R$ matrix on the two
sublattices.  In general, this vertex model is not solvable.

The peculiarity of the antiferromagnetic line is that the vertex
model---and hence the Potts model, at least in principle---becomes
solvable because, when $v=u\pm {\pi\over 2}$, it can be considered as
a particular case of models exhibiting ``$Z$ invariance''
\cite{BaxterI}.  In general, such models are obtained by associating a
spectral parameter $h_{i}$ with every horizontal line (with label
$i$), a spectral parameter $v_{j}$ with every vertical line, such that
the $R$ matrix at the vertex $ij$ is $R\left(h_{i}-v_{j}\right)$. On
the antiferromagnetic line, the $v_{j}$ are alternatingly $0,{\pi\over
  2}$, while the $h_{i}$ are alternatingly $u,u+{\pi\over 2}$. Thanks
to the special shift ${\pi\over 2}$, and the fact that $u$ can be
identified with $u\pm \pi$, there are indeed only two types of
vertices for the resulting vertex model, with interactions given by
$R(u)$ and $R\left(u+{\pi\over 2}\right)$ respectively (see the
appendix for more details).  The isotropic case ($K_{1}=K_{2}$,
$x_{1}=x_{2}$) would correspond to $u={\gamma\over 2}+{\pi\over 4}$.
 
While vertex models with imaginary heterogeneities of the spectral
parameter have been studied in some detail \cite{RS}, the situation of
a real heterogeneity does not seem to have attracted much attention.
We will now see how the vertex model exhibits truly remarkable
properties.

We consider a transfer matrix propagating in the vertical direction.
The basic equations can be obtained using standard techniques---they
appeared intially in the pioneering paper by Baxter \cite{BaxterI}.%
\footnote{Note that some misprints have cropped up in this reference:
  in Eq.~(4.40) there should have been a $t^{-2}$ instead of $t^{2}$ on the
  right-hand side, as confirmed by the analysis of the analytic Bethe
  Ansatz, Eq.~(5.4) of that same paper.}

\subsection{Ground state energy, degeneracies}

We will use modern Quantum Inverse Scattering notations, following
\cite{RS}. We introduce the kernel
\begin{equation}
  \phi(\alpha)=i\ln {\sinh{1\over 2}(\alpha+2i\gamma)\over
    \sinh{1\over 2}(\alpha-2i\gamma)}
\end{equation}
together with the source terms 
\begin{equation}
  p_{1}(\alpha)=i\ln {\sinh{1\over 2}(\alpha-i\gamma)\over
    \sinh{1\over 2}(\alpha+i\gamma)}
\end{equation}
and 
\begin{equation}
  p_{-1}(\alpha)=i\ln {\cosh{1\over 2}(\alpha-i\gamma)\over
    \cosh{1\over 2}(\alpha+i\gamma)}
\end{equation}
Recall that the parameter $\gamma$ is related to $Q$ by $\sqrt{Q}=2\cos\gamma$,
with $\gamma\in [0,{\pi\over 2}]$. The Bethe equations read then
symbolically, for a system of $2N$ vertical lines (this corresponds to
$N$ Potts spins),
\begin{equation}
  \left( e^{ip_{1}(\alpha)}\right)^{N}\left( e^{ip_{-1}(\alpha)}\right)^{N}
  \prod_{\alpha'}e^{i\phi(\alpha-\alpha')}=
  e^{i\psi}
\end{equation}
Here, we have introduced an adjustable phase $\psi$  corresponding to 
twisted boundary conditions. 

The eigenvalue of the transfer matrix has a complicated expression, 
but in the limit of $N$ large to which we will restrict, it reduces to 
\begin{equation}
  \Lambda\propto 
  \prod_{\alpha}e^{ip_{1}(\alpha+2iu)}e^{ip_{-1}(\alpha+2iu)}\label{energ}
\end{equation}
Rather than the vertex model transfer matrix, we will consider  the 
hamiltonian, obtained
in the strongly anisotropic limit $u\rightarrow 0$. Its eigenvalues 
read 
\begin{equation}
  E=c\sum_{\alpha}{d\over d\alpha}\left[p_{1}(\alpha)+p_{-1}(\alpha)\right]
  \label{energi}
\end{equation}
where $c$ is a (positive)  constant to be chosen below. 

A remarkable property of these Bethe equations is that they, like the
energy, are invariant under a global shift of the roots
$\alpha\rightarrow \alpha+i\pi$. This is due to the peculiar form of
the heterogeneities; in general, the invariance is only under
$\alpha\rightarrow \alpha+2i\pi$.  Equivalently, one can think of this
as invariance under $\gamma\rightarrow \pi -\gamma$.  States for which
the root configuration is invariant under $\alpha\rightarrow
\alpha+i\pi$ will correspond to non-degenerate eigenvalues, while
states which are not correspond to eigenvalues degenerate twice.

Writing the contribution of the roots to the energy more explicitly
as
\begin{equation}
  E=c\sum_{\alpha} {2\sin2\gamma\over\cosh 2\alpha-\cos 2\gamma}
\end{equation}
we can conjecture that the ground state is obtained (at least when the
phase $\psi=0$) by filling the lines $\hbox{Im }\alpha={\pi\over 2}$
and $\hbox{Im }\alpha=-{\pi\over 2}$.  The source terms are of a new
type then, which we will denote $q_{1},q_{-1}$. The Fourier transform
(with conventions $\hat{f}(k)=\int_{-\infty}^{\infty}
f(\alpha){\rm e}^{ik\alpha}d\alpha$) is
\begin{equation}
     \widehat{\dot{q}}_{1}+\widehat{\dot{q}}_{-1}=-2\pi{\sinh\gamma 
     k\over \sinh \pi k/2}
\end{equation}
The other kernels to consider are $\phi_{11}\equiv\phi$ and
$\phi_{1,-1}(\alpha)\equiv\phi(\alpha\pm i\pi)$, with Fourier
transforms
\begin{eqnarray}
  \hat{\phi}_{11}&=&-2\pi {\sinh(\pi-2\gamma)k\over\sinh \pi 
    k}\nonumber\\
  \hat{\phi}_{1,-1}&=&2\pi {\sinh 2\gamma k\over\sinh \pi k}
\end{eqnarray}
The minus sign in the source term leads to a ground state where
rapidities are a decreasing function of the Bethe integers, and we
get, in the continuum version
\begin{eqnarray}
  2\pi[\rho_{1}+\rho_{1}^{h}]&=&-(\dot{q_{1}}+\dot{q}_{-1})-\dot{\phi}_{11}
  \star\rho_{1}-
  \dot{\phi}_{1,-1}\star\rho_{-1}\nonumber\\
  2\pi[\rho_{-1}+\rho_{-1}^{h}]&=&-\dot{q_{1}}+\dot{q}_{-1})-\dot{\phi}_{-1,1}
  \star\rho_{1}-
  \dot{\phi}_{-1,-1}\star\rho_{-1}
\end{eqnarray}
Solving these equations gives the densities in the ground state. It is
most useful to go one step further and write ``physical equations''
where on the right-hand side appear holes in the ground state
distributions, that is densities of excitations over the physical
vacuum:
\begin{eqnarray}
  2\pi(\rho_{1}+\rho_{1}^{h})&=&s(\alpha)+\Phi_{11}\star\rho_{1}^{h}+
  \Phi_{1,-1}\star\rho_{-1}^{h}\nonumber\\
  2\pi(\rho_{-1}+\rho_{-1}^{h})&=&s(\alpha)+\Phi_{-1,1}\star\rho_{1}^{h}+
  \Phi_{-1,-1}\star\rho_{-1}^{h}
\end{eqnarray}
The kernels here are defined by their Fourier transforms
\begin{eqnarray}
  \hat{\Phi}_{11}&=&\hat{\Phi}_{-1,-1}=-2\pi {\cosh(\pi-3\gamma)k\over 
    2\sinh \gamma k\sinh(\pi-2\gamma)k}\nonumber\\
  \hat{\Phi}_{1,-1}&=&\hat{\Phi}_{-1,1}=2\pi {\cosh\gamma k\over 
    2\sinh\gamma k\sinh(\pi-2\gamma)k}
\end{eqnarray}
The function $\hat{s}={\pi\over \cosh(\pi-2\gamma)k/2}$,
$s(\alpha)={\pi\over (\pi-2\gamma) \cosh{\pi\over
    \pi-2\gamma}\alpha}$.  From this the ground state densities first
follow, $\rho_{1}^{\rm gs}=\rho_{-1}^{\rm gs}={1\over 2\pi} s(\alpha)$. The
ground state energy per vertex is then
\begin{equation}
  {E^{\rm gs}\over N}=-c\int {\sinh (\gamma k)\over\sinh(\pi 
    k/2)\cosh(\pi-2\gamma)k/2}\label{energgs}
\end{equation}
It is interesting to compare  this result with Baxter's expressions 
for the free energy \cite{BaxterII}.
Using analyticity techniques, he finds that the free energy per site
can be written as $f=f(K_{1})+f(K_{2})\equiv f(u)+f(u+\pi/2)$ where
the function $f$ (denoted by $\phi$ in his work) is given in
Eqs.~(30a)--(30d) of \cite{BaxterII}. To compare with our hamiltonian
we need to let $u$ approach $0$ from below, and take a derivative, so
we must consider the expression
\begin{eqnarray}
  \frac{d}{du}\left[f(u)+f(\pi/2-u)\right]&=&
  \frac{d}{du}\left[-\int dk{\sinh^{2}(\gamma k)\sinh (2uk)\over k 
      \sinh(\pi k)\sinh(\pi-2\gamma)k}+
    \int dk{\sinh(\gamma k)\sinh(\pi-\gamma)k\sinh (2uk)\over k\sinh(\pi ki)
      \sinh(\pi-2\gamma)k}\right]\nonumber\\
  &=&
  2\int dk {\sinh (\gamma k)\over \sinh(\pi 
    k/2)\cosh(\pi-2\gamma)k/2}\label{ffirstdomain}
\end{eqnarray}
This agrees with our ground state energy, given in
(\ref{energgs}), and constitutes an interesting check of our
equations.

\subsection{Excitations}

The ``physical'' Bethe equations immediately show that there are two
branches of massless excitations (at least) obtained by making holes
in the ground state at $\pm\infty$ on the lines $\hbox{Im }\alpha=
{\pi\over 2}$ and $\hbox{Im }\alpha=-{\pi\over 2}$. This is {\sl
  twice} the usual result for the XXZ chain, and leads one to expect
that the central charge will be equal to two, a result we will
confirm below. The rapidity of the excitations (such that $e=\pm
p\propto {\rm e}^{\theta}$) is $\theta= {\pi\over \pi-2\gamma}\, \alpha$.  As
for the energy, one has
\begin{equation}
  {E\over N}={E^{\rm gs}\over N}+ c\int 
  s(\alpha)\left[\rho^{h}_{1}(\alpha)+\rho_{-1}^{h}(\alpha)\right]
\end{equation}
so relativistic invariance requires $c={(\pi-2\gamma)\over \pi}$.
  
The kernels $\Phi$ describe a complex scattering theory, which we do
not yet entirely understand. A crucial fact is that these kernels
diverge at small $k$, a very unusual feature, that is also encountered
in the studies of models based on super algebras \cite{EssFSal}. We
will argue below that it implies the existence of a continuum
component of the spectrum.  One can however observe some drastic
simplifications for ``symmmetric excitations'', i.e., excitations which
are identical for the two lines $\hbox{Im }\alpha= {\pi\over 2}$ and
$\hbox{Im }\alpha=-{\pi\over 2}$. For these, the ``effective kernel''
is obtained as $\Phi=\Phi_{11}+\Phi_{1,-1}$,
\begin{equation}
  \widehat{\Phi}\equiv -2\pi{\sinh (\pi-4\gamma)k/2\over 2\sinh \gamma k
    \cosh (\pi-2\gamma)k/2}
\end{equation}
This is the soliton-soliton kernel for the sine-Gordon model regime
with ${\beta^{2}\over 8\pi}={2\gamma\over\pi}$.  This observation has
a simple origin. Consider again the Bethe equations: it is clear that
there is a particular type of solutions obtained when the real parts
of the roots on $\hbox{Im }\alpha={\pi\over 2}$ and $\hbox{Im
}\alpha=-{\pi\over 2}$ coincide. The equations for these real parts
are then of the form
\begin{equation}
  \left({\sinh 
      (\alpha-i\gamma+i\pi/2)\over\sinh(\alpha+i\gamma-i\pi/2)}\right)^{N}=e^{i\psi}
  \prod_{\alpha'}{\sinh (\alpha-\alpha'-2i\gamma)\over
    \sinh (\alpha-\alpha'+2i\gamma)}\label{bethesymm}
\label{XXZ_BAE}
\end{equation}
while the energy reads 
\begin{equation}
  E=-{2(\pi-2\gamma)\over \pi} \sum_{\alpha} {2\sin 2\gamma\over \cosh 
    2\alpha+\cos 2\gamma}
\end{equation}
Comparing with the results for the XXZ chain (with relativistically
invariant normalization)
\begin{equation}
  H=-{\gamma'\over2\pi\sin\gamma'}\sum_{i} \left[
    \sigma_{i}^{x}\sigma_{i+1}^{x}+\sigma_{i}^{y}\sigma_{i+1}^{y}-
    \cos\gamma'
    \sigma_{i}^{z}\sigma_{i+1}^{z} \right] \label{XXZ}
\end{equation}
we see from standard works on this topic \cite{Alcaraz} that this
spectrum will lead to {\sl twice} the central charge and critical
exponents of the XXZ chain (\ref{XXZ}) with parameter
$\gamma'=\pi-2\gamma$.  On the other hand it is known that the
conformal field theory limit of this XXZ chain is a free boson with
coupling constant $g={2\gamma\over\pi}$ (in agreement with the value
${\beta^{2}\over 8\pi}={2\gamma\over\pi}$ obtained before for the
scattering matrix). Notice in particular that the limit $Q\rightarrow
0$ corresponds to $\gamma'=0$, i.e., the XXX chain. In other words,
there has to be a whole class of excitations in our system for which
the gaps will be the same as the gaps of this XXZ chain (\ref{XXZ}),
{\sl up to a factor two} (this should hold with or without twist angle
$\psi$). We call this `the XXZ subset' of our spectrum.  It is
tempting to speculate that it corresponds exactly to the
configurations of roots which are invariant under the extra $Z_{2}$
symmetry $\alpha\rightarrow\alpha+i\pi$.  Of course, there has to be
many more excitations, since doubling the spectrum of the XXZ chain
creates many gaps in the conformal towers.  These excitations should
appear with an extra degeneracy of two, since they corresponds to
configurations of roots which are not symmetric.
 
\subsection{A quick analysis of finite-size effects}
 
We can make further progress on this question by analyzing in more
details the finite-size spectrum from the Bethe Ansatz.  For this, we
go back to the physical Bethe equations involving holes in the
distributions of strings and antistrings. It is interesting to
determine the dressed charges of the excitations. Normally, this is an
easy task. Write the bare equations as
$\hat{\rho}+\hat{\rho}^{h}=\hat{p}+K\hat{\rho}$. It follows that $
\hat{\rho}+\hat{\rho}^{h}={\cal Z}\hat{p}+(1-{\cal Z})\hat{\rho}^{h}$,
where ${\cal Z}=\left(1-K\right)^{-1}$.  Standard formulae then exist to
express the gaps in terms of the matrix ${\cal Z}$ \cite{deVega}:
\begin{equation}
  h+\bar{h}= \Delta n^{t}{\cal Z}^{-2}\Delta n+d^{t}{\cal Z}^{2}d
  \label{hbarh}
\end{equation}
where $\Delta n$ is the two-column vector $\left(\begin{array}{c}
    \Delta n_{1}\\
    \Delta n_{2}
 \end{array}\right)$, and $\Delta n_{i}$ the change of solutions of the 
$i^{th}$ type. Similarly, $d$ is the two-column vector
$\left(\begin{array}{c}
    d_{1}\\
    d_{2}
  \end{array}\right)$, and $d_i$ the number of solutions of the $i^{th}$ type
backscattered from the left to the right of the Fermi sea. In (\ref{hbarh}),
$t$ denotes transposition.

Here however we encounter a major snag: the matrix ${\cal Z}$ has all its
elements infinite, as the kernels diverge when $k\rightarrow 0$. We
have taken the most naive approach possible, and assumed that the
standard formulas still made sense in this case, after maybe some
regularization. It turns out that such regularization is easy: just
keep $k$ small but non-zero in the matrix elements. One can then
calculate the inverse of ${\cal Z}$, and let $k\rightarrow 0$ in the end.
One finds
\begin{equation}
  {\cal Z}^{-1}\equiv \lim_{k\rightarrow 0}{1\over 
   {\cal Z}(k)}={\gamma\over\pi}\left(\begin{array}{cc} 
      1/ 2&1/2\\
      1/2&1/ 2\end{array}\right)
\end{equation}
and thus 
\begin{equation}
  \Delta+\bar{\Delta}={\gamma\over 2\pi} (\Delta n_{1}+\Delta n_{2})^{2}+0\times (\Delta n_{1}-\Delta 
  n_{2})^{2}+{\pi\over 8\gamma}(d_{1}+d_{2})^{2}+\infty \times 
  (d_{1}-d_{2})^{2}\label{spectrum}
\end{equation}
We recover here the XXZ subset (see later) for $d_{1}=d_{2}$. On top
of this however, we have a continuum of soft modes corresponding to
$\Delta n_{1}\neq \Delta n_{2}$. This is similar to observations in
\cite{JacSal05} and \cite{EssFSal} where the soft modes were
interpreted as arising from a non-compact boson. We therefore believe
that the continuum limit of the untwisted vertex model is described by
a compact boson (called $\phi_{1}$ in what follows) with $g={1\over
  t}$ {\sl plus} a non-compact boson (called $\phi_{2}$) giving a
continuum of soft modes on top of the gaps for the first boson.
 
While the action of the twist angle $\psi$ on $\phi_{1}$ (that is, the
XXZ subset of the spectrum) is well understood, the action on
$\phi_{2}$ seems to be more subtle, and cannot be captured by our
simple analysis. We shall see that even the parity of $N$ affects
$\phi_{2}$ in a non-trivial way.

\subsection{Some remarks on extra symmetries}
 
\begin{figure}
  \centering
  \includegraphics[scale=0.5]{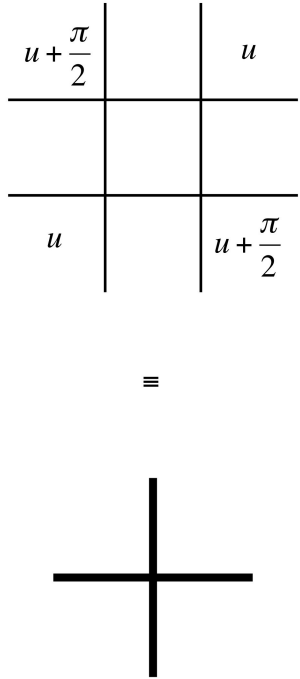}
  \caption{A block of four vertices can be considered as a 
    ``supervertex'' for a model with $2\times 2=4$ degrees of freedom 
    per edge. 
  }
  \label{block}
\end{figure}

The Bethe equations expressed in terms of roots at $\hbox{Im
}\alpha=\pm {\pi\over 2}$ look like equations for a rank two algebra,
and suggest the existence of two $U(1)$ conserved quantities, that is,
an extra quantity on top of the spin $S^{z}$. One can get an idea of
what this extra quantity might be by considering the limit
$Q\rightarrow 0$. We observed in \cite{JacSal05} that in this limit,
the Bethe equations and the energy of the model coincide with those of
the integrable $osp(2/2)$ spin chain in the fundamental
representation. One can in fact go further (see the appendix), and
prove that the $osp(2/2)$ integrable $R$ matrix can be obtained by
considering a ``block'' made up of four elementary vertices of the 6
vertex model. In this case, the extra $U(1)$ charge is the $B$ quantum
number of the $osp(2/2)$ algebra; this is defined initially for a pair
of neighbouring spins $|S_{2i-1}S_{2i}\rangle$ by $B=0$ for $|++\rangle$
or $|--\rangle$, and $B=\pm 1/2$ for $|+- \rangle \pm i|-+ \rangle$
(see Fig.~\ref{block}), and then extended to the whole system by
summing over $i$. It is very interesting to study what becomes of
this extra symmetry on the whole antiferromagnetic line, and the
relation with the $osp_{q}(2/2)$ integrable $R$ matrix---we will report
about this elsewhere.  In any case, the quantity $\Delta n_{1}-\Delta
n_{2}$ plays the role of an extra $U(1)$ charge along the whole
antiferromagnetic line.

On top of this charge, the antiferromagnetic line exhibits an extra
$Z_{2}$ symmetry, as already mentioned in \cite{SaleurI}.  This can be
seen most clearly by considering again the block of spins in Fig.~\ref{block}
and writing its $\check{R}$ matrix algebraically as (for
notations see \cite{SaleurI})
\begin{equation}
  \check{R}=X_{2}(u)X_{1}(v)X_{3}(v)X_{2}(u),~~~v=u\pm {\pi\over 2}
\end{equation}
Using the Yang-Baxter equation repeatedly, one can show that the
operator
\begin{equation}
  C=X_{1}\left({\pi\over 2}\right)X_{3}\left({\pi\over 2}\right)
\end{equation}
commutes with $\check{R}$, where the fact that $X\left({\pi\over
    2}\right)=X\left(-{\pi\over 2}\right)$ is crucial. On the other
hand, one has \cite{SaleurI}
\begin{equation}
     X\left(\pm {\pi\over 2}\right)=P_{1}-P_{0}
\end{equation}
where $P_{1},P_{0}$ are the projectors on (quantum) spin one and zero
representations in the product of two spin $1/2$ representations, so
$C^{2}=1$.
  
We can now discuss briefly the phase diagram of the Potts model in
terms of the symmetries of the associated vertex model.  The selfdual
case $x_{1}={1\over x_{2}}$ gives rise to two kinds of physical
behaviors, which correspond in the isotropic case to two selfdual
lines, $x_{1}=x_{2}=1$ and $x_{1}=x_{2}=-1$. In the selfdual case,
the symmetry of the associated vertex model is $U(1)\times Z_{2}$. The
$U(1)$ is just the conservation of the spin $S^{z}$, while the $Z_{2}$
is the symmetry of translation by one lattice site. The latter
symmetry, in the continuum limit, becomes another $U(1)$
\cite{Affleck} so the symmetry of the continuum limit on the selfdual
lines is $U(1)\times U(1)$: this corresponds to chiral symmetries for
left and right fermions, or symmmetry
$\phi_{L}\rightarrow\phi_{L}+\alpha$,
$\phi_{R}\rightarrow\phi_{R}+\beta$, of the free bosonic continuum
limits. Going away from the selfdual lines breaks the $Z_{2}$. If one
starts from the physical critical line ($x_{1}=x_{2}=1$), the model
becomes massive. If one starts from the non-physical critical line
($x_1=x_2=-1$), one enters the Berker-Kadanoff phase \cite{SaleurI} where the
$Z_{2}$ symmetry breaking operator is irrelevant, and the full $U(1)$
is actually restored in the continuum limit.

The antiferromagnetic critical line marks the boundary of the
Berker-Kadanoff phase.  As just discussed, on this line an extra
$U(1)\times Z_{2}$ symmetry appears, and it is reasonable to
expect that in the continuum limit, this also gets enhanced into the
full $U(1)\times U(1)$, corresponding to another free boson. This is
in agreement with the Bethe Ansatz calculation of the central charge.
However, when $Q=0$ it has been argued in \cite{JacSal05} that this
boson is non-compact.  This feature must be true along the full
antiferromagnetic line in fact, as follows from the finite-size
corrections to the gaps.

\subsection{The spectrum of the untwisted vertex model for $N$ even}

One subtlety is that results for $N$ odd and $N$ even turn out in fact
to be profoundly different. We do not understand this fully, but
notice this can be expected, as the foregoing analysis applied mostly
to $N$ even, which allowed one to treat the two lines
$\hbox{Im}\alpha=\pm {\pi\over 2}$.  The case of $N$ odd will be
considered later. In what follows, we parametrize $\gamma={\pi\over
  t}$, with $t\in [2,\infty]$.

To start, we consider the XXZ subset some more. Magnetic gaps read
\begin{equation}
  \Delta=2 \times {g(\tilde{S}^{z})^{2}\over 4}={(\tilde{S}^{z})^{2}\over t}
\end{equation}
where $\tilde{S}^{z}$ is the spin of the ``equivalent XXZ
chain''---that is the chain for which the Bethe equations are
(\ref{bethesymm}).  Since the antiferromagnetic system has $2N$ spins,
whereas the XXZ chain described by (\ref{XXZ_BAE}) is for $N$ spins only,
we must rescale the spin according to
$\tilde{S}^{z}={S^{z}\over 2}$, where $S^{z}$ is the spin in our
initial system, i.e., $S^{z}=\Delta n_{1}+\Delta n_{2}$ in
(\ref{spectrum}). Therefore
\begin{equation}
  \Delta={(S^{z})^2\over 4t}\label{magneticgap}
\end{equation}
where $S^{z}$ (the spin in our initial system) is an integer since we
have a even number of spins.

One can see here again why parity effects are likely to occur.  If $N$
is even, since we must also have an even number of upturned spins to
make up pairs of Bethe roots in the ``folding'' of the equations, the
result strictly speaking will only apply to $S^{z}$ even, including
the ground state. Numerical study shows that in fact it holds for
$S^{z}$ odd as well. If $N$ is odd on the other hand, since we must
also have an even number of upturned spins to make up pairs of Bethe
roots in the ``folding'' of the equations, the result strictly
speaking will only apply to $S^{z}$ odd. Numerically, we have observed
indeed that for $S^{z}$ even, including the ground state $S^{z}=0$,
results for $N$ odd obey a different pattern.  See section 7 below.
 
The electric charges in the $S^{z}=0$ sector of the equivalent XXZ
chain are integer ($e=d_{1}=d_{2}$), and give rise to the following
gaps $\Delta=2\times {e^{2}\over 4g}={te^{2}\over 4}$. Combining all
the excitations, we thus get exponents from a {\sl single} free boson,
$\Delta_{em}={1\over 4}\left({m\over\sqrt{t}}\pm
  e\sqrt{t}\right)^{2}$, with $e$ and $m$ integers.\footnote{In
  general, we will denote by $\Delta$ the (rescaled) gaps with respect
  to the central charge $c=2$, and reserve the notation $h$ for the
  (rescaled) gaps with respect to the true ground state of the
  antiferromagnetic Potts model.}
 
Meanwhile, the non-compact boson adds one unit to the central
charge, and `decorates' these gaps by soft modes, which would manifest
theshelves in a numerical study as very large degeneracies
\cite{EssFSal}. One can write this in a compact form by considering
the generating function of levels
\begin{equation}
  Z_{\rm conj}=\sum_{\Delta,
    \bar{\Delta}} q^{\Delta-c/24}\bar{q}^{\bar{\Delta}-c/24}=
  \int d\alpha (q\bar{q})^{\alpha^{2}-1/24}\times 
  {1\over 
    \eta\bar{\eta}}\sum_{e,m}q^{\Delta_{em}}\bar{q}^{\bar{\Delta}_{em}}
\end{equation}
with $\Delta_{em}(\bar{\Delta}_{em})={1\over 4}\left({m\over
    \sqrt{t}}\pm e\sqrt{t}\right)^{2}$. It would coincide with the 
    partition function of the doubly 
    periodic vertex model with $q={\rm e}^{-2 \pi M/N}$
the modular parameter for a system of size $N \times M$ and $\eta =
q^{1/24} \prod_{n=1}^\infty (1-q^n)$ Dedekind's eta function.

This means that in the sector of vanishing magnetization, one should
observe a continuum spetrum above the ground state. In the sector with
magnetization unity, one should observe a continuous spectrum over the
basic magnetic gap, etc. Numerical study by direct diagonalization of
the transfer matrices (possible for $N \le 12$) is compatible with
this result, though the sizes accessible are too small to provide a
complete confirmation. In \cite{EssFSal} where the Bethe equations
themshelves were studied for a simimlar problem, it was found that
sizes as large as $N \simeq 10^4$ were necessary.

  \section{The spectrum of the twisted vertex model and the  Potts model 
  transfer matrices 
  on the antiferromagnetic line}
  
The usual strategy is, once the continuum limit of the untwisted 
vertex model is identified, to evaluate the effect of the twisting and 
other ingredients involved in the correspondence with the loop/cluster 
formulation, and gain in this way knowledge of the conformal 
properties of the Potts model itself. The process in the 
antiferromagnetic region is however more involved than in the 
ferromagnetic region. To start, we discuss the Berker-Kadanoff phase 
in more details.

\subsection{The Berker-Kadanoff phase}
 
Recall first that the untwisted vertex model in the whole
Berker-Kadanoff (BK) phase is described in the continuum limit, like
on the non-physical selfdual critical line, by a single free boson
with $g={1\over t}$. It is important to notice that this free boson is
in fact exactly the same as the one describing the XXZ subset on the
antiferromagnetic critical line.
 
The first step in going from the vertex model to the loop/cluster
representation of the Potts model is to introduce a twist in the
vertex model. We have explained that this
twist has a convenient geometrical interpretation
if one decomposes the vertex configurations in loops: without it,
loops which are non-contractible around the cylinder get a weight $2$,
instead of the required $\sqrt{Q}$ for the loop representation of the
Potts model.  Suppose we want to give to non-contractible loops  a
more general weight, $w=2\cos\pi \alpha$. We introduce a seam or twist
in the vertex model and the associated conformal weight should give
the effective central charge
\begin{equation}
  c=1\times \left(1-{6\alpha^{2}\over g}\right)=1-6t \alpha^{2}
\end{equation}
Since non-contractible loops come in pairs, we have $\alpha$ defined
modulo an integer: these are the allowed values of the electric charge
in the Coulomb gas. The $Q$-state Potts model itself corresponds to
$\alpha\equiv e_{0}={1\over t}$, leading, in principle, to the central
charge $c=1-{6\over t}$.
  
This latter value is definitely the central charge of the twisted
vertex model. But it is not the central charge of the loop/cluster
Potts model!
 
The point is that the eigenvalues of the loop or cluster model are in
general (as discussed in section 2) a {\sl subset} of the eigenvalues of the vertex model,
essentially those associated with ``highest weights'' only. The
determination of this subset is based on algebraic considerations, and
follows some simple rules \cite{PS}.  For instance, eigenvalues of the
vertex model transfer matrix in the sector of spin $S^{z}=1$ and with
no twist are also exactly observed in the sector $S^{z}=0$ and with a
twist $\alpha={1\over t}$ (i.e., giving to non-contractible loops the
weight $\sqrt{Q}$, that is $\alpha=e_{0}$). Such a ``descendent''
eigenvalue is thrown away in the loop or cluster formulation.
    
\begin{figure}
  \centering
  \includegraphics[width=4in]{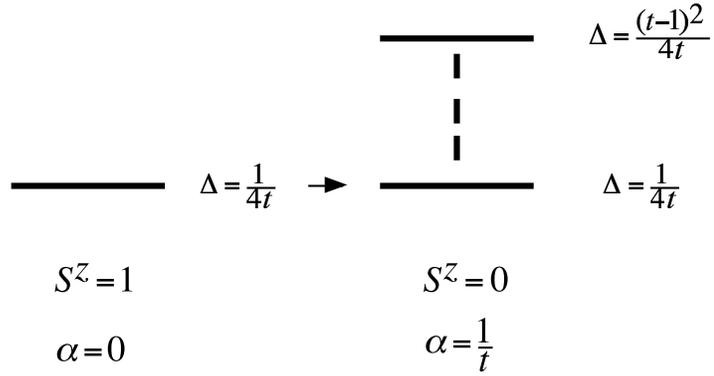}
  \caption{A schematic representation of  the behavior of some  
    levels  (in finite size)  of the vertex model in the 
    BK phase, for $t$ generic. Levels which are at the head of 
    an arrow are thrown away in going from the vertex to the loop 
    or cluster formulation. The level $\Delta={1\over 4t}$ is thus 
    always discarded, leaving generically the next level 
    $\Delta=(t-1)^{2}/4t$ to determine the central charge 
    $c=1-6(t-1)^{2}/t$ for the loop or cluster model  for $t$ generic.
  }
  \label{BKI}
\end{figure}

In the BK phase, the dimension associated with electric charge
$e_{0}={1\over t}$, since it is the same as the dimension coming from
the sector with no electric charge (twist) and $S^{z}=1$, and equal to
$\Delta_{e_0,0}={1\over 4t}$, is thrown away in the truncation from
vertex to loop (or cluster). This means that the central charge of the
twisted vertex model and the loop/cluster model are different in this
phase. For the vertex model one has $c=1-{6\over t}$ while the central
charge of the loop or cluster model follows from the next choice,
$e_{0}=1-{1\over t}$, leading instead to central charge
$c=1-{6(t-1)^{2}\over t}$, all this for $t$ generic: this is
illustrated in Fig.~\ref{BKI}.

\begin{figure}
  \centering
  \includegraphics[scale=.4,angle=-90 ]{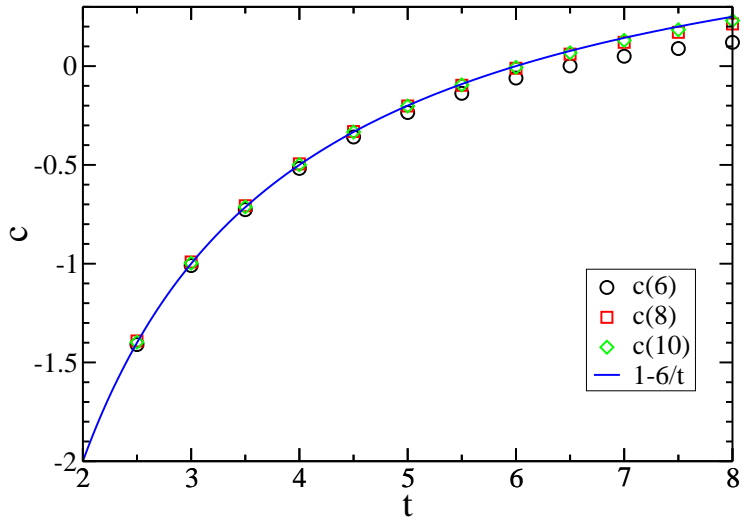}
  \caption{Numerical estimates of the 
    central charge for the twisted vertex model on the non-physical 
    selfdual line. The continuous curve is the expected value, 
    $c=1-6/t$. Diagonal geometry.
  }
  \label{num1}
\end{figure}

\begin{figure}
  \centering
  \includegraphics[scale=.4,angle=-90]{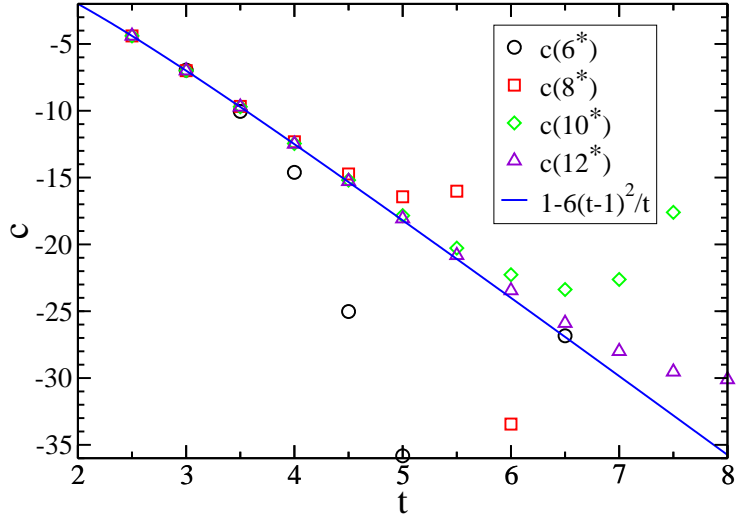}
  \caption{Numerical estimates of the
    central charge for the loop model on the non-physical 
    selfdual line. The continuous curve is the expected value, 
    $c=1-6(t-1)^{2}/t$. Axial geometry.
  }
  \label{num2}
\end{figure}

We have performed careful numerical checks of this phenomenon, with
some results presented in Figs.~\ref{num1}--\ref{num2}. A technical
note here: one can essentially write two transfer matrices, one
propagating along the axial direction of the original Potts lattice
(and thus diagonally for the medial lattice, which is the lattice
where the loops or the vertex spins live), and one propagating
diagonally for the Potts lattice (and thus axially for the loops and
vertex spins). Results will not (except for parity effects) depend on
the geometry in the thermodynamic limit, but might converge better in
finite size with one choice or the other. Note that the first
formulation is better suited for using the Temperley-Lieb algebra and
quantum group symmetries. It is the one we use most. In all figure
captions we will use the name diagonal and axial in reference to the
medial lattice (see also footnotes~\ref{fnTM1}--\ref{fnTM2}).

\begin{figure}
  \centering
  \includegraphics[width=4in]{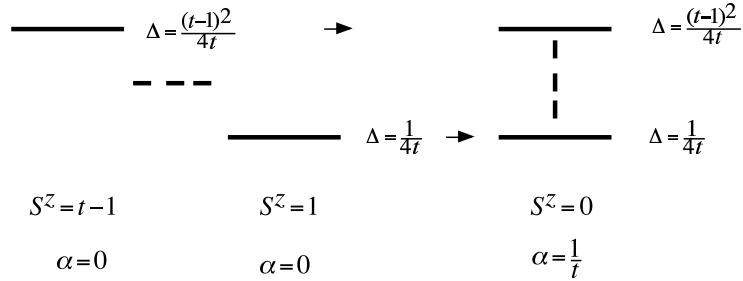}
  \caption{A schematic representation of the behavior of some levels
    (in finite size) of the vertex model in the BK phase, for $t$ an
    integer. Now the two levels indicated on the right are both
    discarded. Many more levels are in fact discarded, leading to the
    singularities at the Beraha numbers within the BK phase.  }
  \label{BK}
\end{figure}

Notice that for $t$ an integer (and only then), the eigenvalue
associated with $e_{0}=1-{1\over t}$ is also thrown away because it is
a descendent from the sector with no twist and spin $S^{z}=t-1$, as
illlustrated in Fig.~\ref{BK}. This is the mechanism that, in the
periodic case, produces the singularities of the BK phase at the Beraha
numbers $Q=4\cos^{2}{\pi\over t}$, with $t$ integer. It is the equivalent
of the slightly simpler mechanism studied in \cite{SaleurI} in the
case of open boundary conditions. In fact, many more levels are
thrown away, so many that even the free energy per unit site of the
(twisted) vertex and loop or cluster models are different at the
Beraha values in the BK phase. The difference of these free energies
vanishes as the antiferromagnetic line is approached. Right on the
antiferromagnetic line, the free energy does not show any singularity
while one crosses a Beraha number (see section 8, in particular
Figs.~\ref{num14}--\ref{num14a}).
	
A note of caution on the case $t$ integer is in order. Of course, the
spectrum of the loop/cluster transfer matrix does not show any
particular behavior when one crosses these values. What happens
however is that the partition function of the model on a torus, say,
would be expressed by weighing these eigenvalues with complicated
non-integer `degeneracy factors', and some of these factors can vanish
when $t$ is integer, a feature we refer to as a level being truncated,
or discarded. Hence what one calls an exponent of the loop/vertex
model has to be defined with great care!

\subsection{Central charge on the antiferromagnetic critical line}

We can now carry out a similar analysis for the model on the
antiferromagnetic critical line. To give non-contractible loops a
weight $w=2\cos\pi\alpha$, we introduce a seam, and the effective
central charge follows immediately as
\begin{equation}
  c=2\times \left(1-{6\alpha^{2}\over 
      g}\right)=2-6t\alpha^{2}\label{twiscen}
\end{equation}
since now $g=\frac{2 \gamma}{\pi} = \frac{2}{t}$ as in section 3.2.
For the $Q$-state Potts model itself we have $\alpha=e_{0}={1\over 
t}$, leading to a central charge $c=2-{6\over t}$. 

\begin{figure}
  \centering
  \includegraphics[scale=.4,angle=-90]{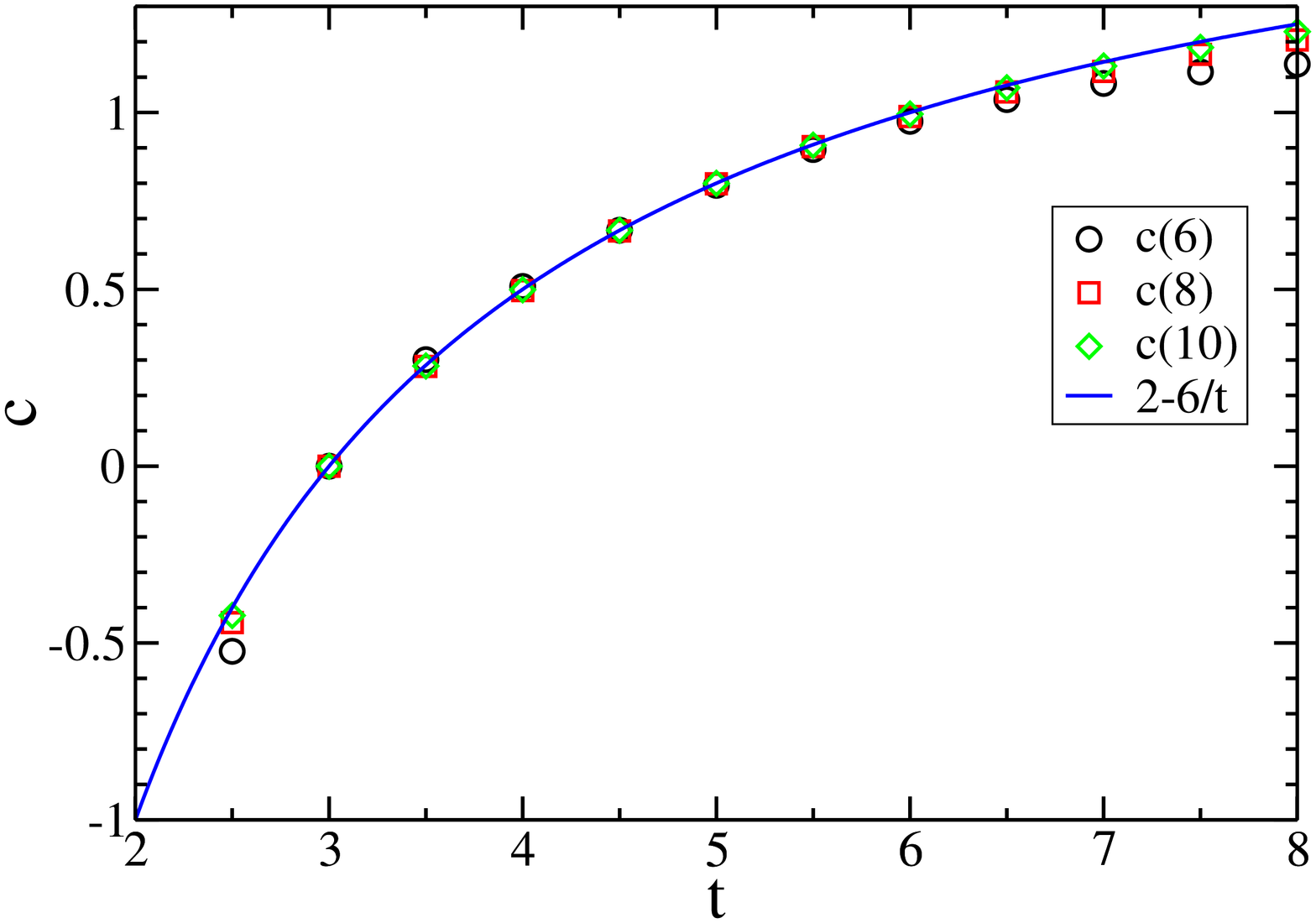}
  \caption{Numerical estimates of the central charge for the twisted
    vertex model on the antiferromagnetic critical line. The
    continuous curve is the expected value, $c=2-6/t$. Diagonal
    geometry.}
  \label{num3}
\end{figure}

\begin{figure}
  \centering
  \includegraphics[scale=.4,angle=-90]{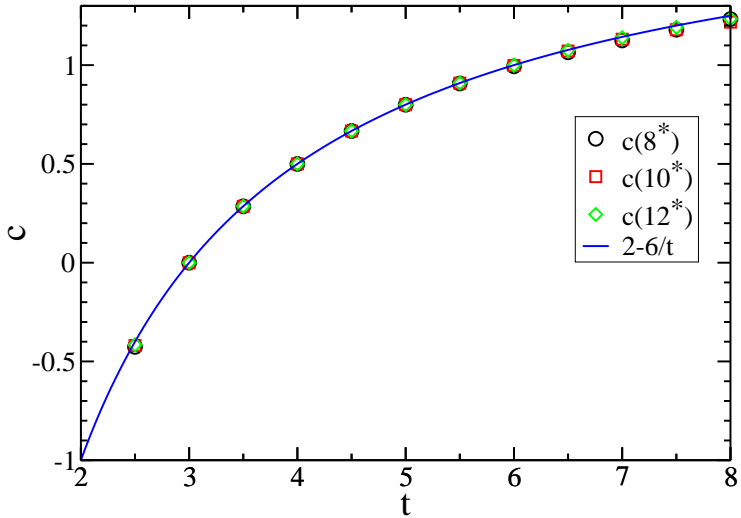}
  \caption{Numerical estimates of the central charge for the loop
    model on the antiferromagnetic critical line. The continuous curve
    is the expected value, $c=2-6/t$. Axial geometry.}
  \label{num4}
\end{figure}

Remarkably, this is not only the central charge of the twisted vertex
model, but {\sl also} the central charge of the loop/cluster Potts
model on this line, as was established earlier \cite{SaleurI}, and can
be checked numerically to great accuracy.  Some results are presented
in Figs.~\ref{num3}--\ref{num4}.  For Fig.~\ref{num3} the transfer
matrix propagates along the diagonal of the medial lattice, for
Fig.~\ref{num4} along the medial lattice itself. Similar results are
obtained for other geometries.
	
This result is remarkable, because we have seen that the central 
charge of the twisted vertex model comes entirely from the XXZ 
sector, and that this sector is described by the same free boson 
as in the Berker-Kadanoff phase. Hence one should have expected the 
level corresponding to (\ref{twiscen}) to be truncated when going from 
the vertex to the loop/cluster formulation, and the central charge of 
the Potts model to be given by another value. What happens is 
the following. 

\begin{figure}
  \centering
  \includegraphics[scale=.4,angle=-90]{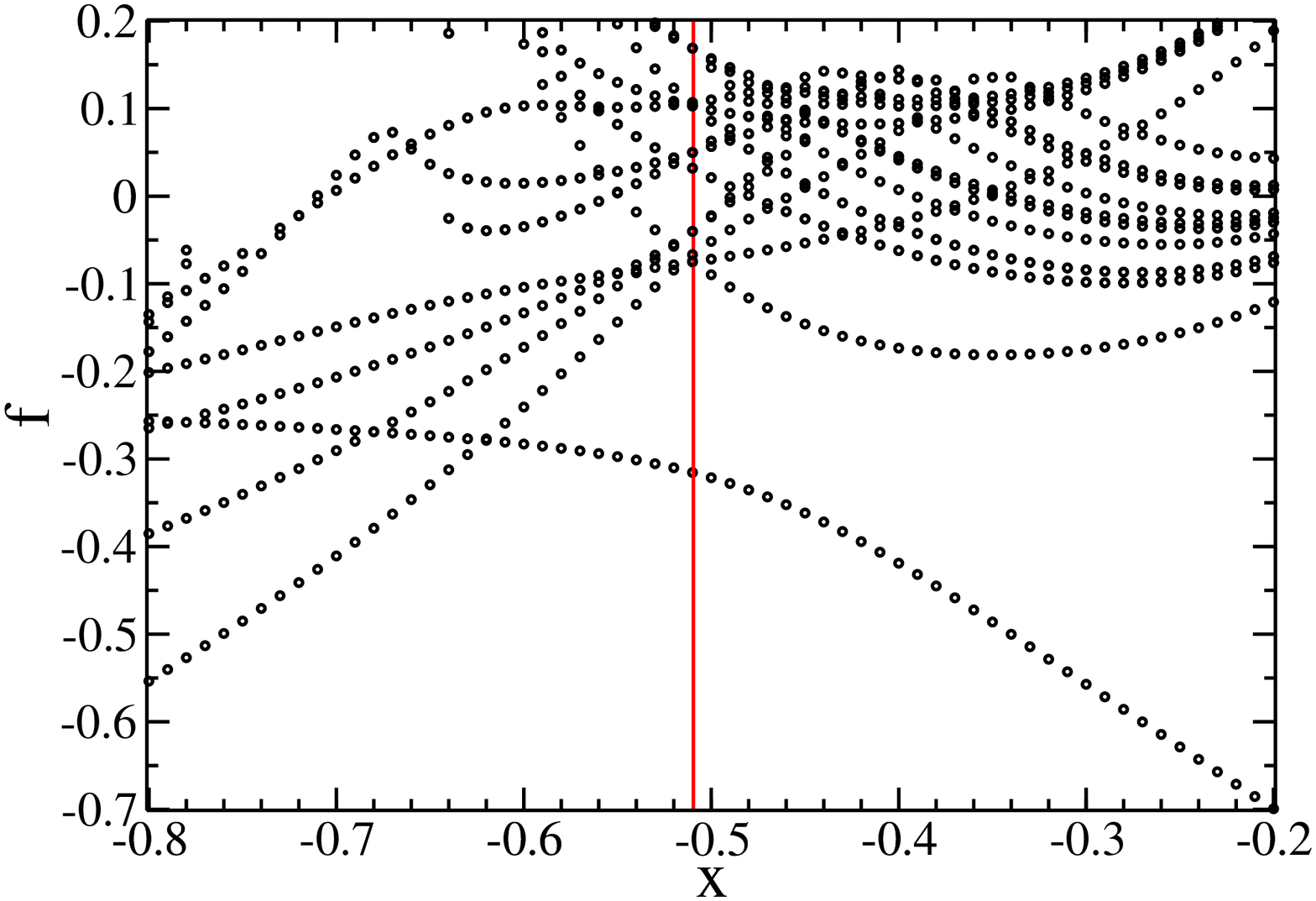}
  \caption{First scaling levels, normalized as free energy densities,
    of the cluster model with $t=5$ and size $N=6$ (diagonal
    geometry). The abscissa is the temperature parameter
    $x=Q^{-1/2}({\rm e}^K-1)$. The vertical line shows the bulk
    antiferromagnetic transition temperature. The ground state on one
    side of the transition becomes a high excitation on the other
    side, and vice versa.}
  \label{num9}
\end{figure}

\begin{figure}
  \centering
  \includegraphics[width=4in]{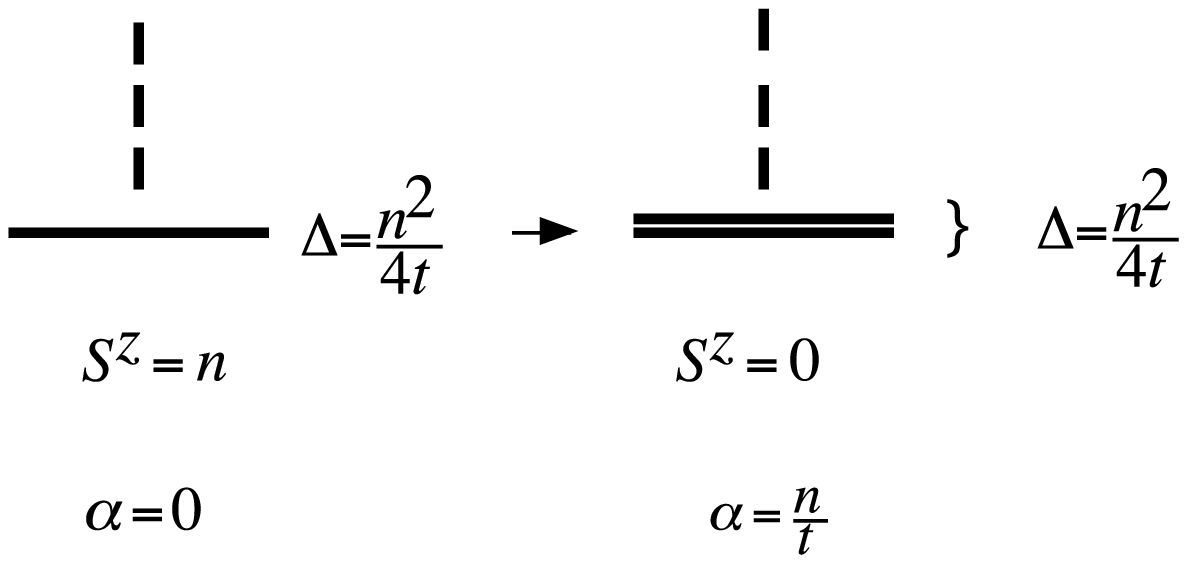}
  \caption{The levels corresponding to Fig.~\ref{BKI}, but on the
    antiferromagnetic line.  The ground state in the sector
    $S^{z}=0$, $\alpha={n\over t}$ has now double degeneracy, so that one
    of the two levels remains after the descendent levels are
    discarded, leading to the central charge $c=2-{6\over t}$, for any
    $t$. The additional level comes down from the very excited part of
    the spectrum as one moves within the BK phase towards the
    antiferromagnetic critical line.}
  \label{BKII}
\end{figure}

Observe that the XXZ subset of the spectrum (in the scaling limit) on
the antiferromagnetic critical line coincides with the spectrum (in
the scaling limit) in the BK phase, with or without twist angle (and
up to an additional contribution of one to the central charge). Since,
because of the doubling effect discussed previously, the vertex model
on the antiferromagnetic critical line has {\sl more} scaling gaps
than in the BK phase, it means that some of the levels which were at
finite distance from the ground state in the BK phase have to cross
and become scaling levels on the antiferromagnetic critical line.  (A
similar crossing phenomenon is clearly visible on the finite-size
levels of the loop/cluster model, see Fig.~\ref{num9}).  One of these
levels happens to merge (asymptotically) with the descendent level, so
after truncation one of them remains, to give the central charge
$c=2-6/t$ still. This is illustrated schematically in Fig.~\ref{BKII}.

This is confirmed numerically: we find that the watermelon operator
$h_{2}=0$, that is the largest eigenvalue of the loop transfer matrix
in the sector with zero legs, and weight $\sqrt{Q}$ for non-contractible
loops is asymptotically degenerate\footnote{More
  precisely, the logarithm of their ratio vanishes faster than $1/N$.}
with the largest eigenvalue in the sector with two legs, and weight
$2$ for non-contractible loops.  Since the latter eigenvalue is also
observed (as a ``descendent eigenvalue'') in the vertex model in the
sector with $S^{z}=0$ and twist angle $\alpha=1/t$ (giving to
non-contractible loops the weight $\sqrt{Q}$), it follows that the latter
sector has a ground state degenerate at least {\sl twice}. This is the
hall mark of a first-order phase transition, and we conclude that {\sl
  the entire antiferromagnetic critical line is in fact a line of
  first-order critical points for the twisted vertex model}. This is
also true of the loop model, even though it has fewer eigenvalues,
because of the result that $h_{2}=0$.%
\footnote{The existence of a first-order critical point in the loop/cluster
formulation was made initially for the $Q\to 0$ limit in Ref.~\cite{Sokal}.}
Finally, it is also true for the
untwisted vertex model (since its ground state is in fact infinitely
degenerate due to the continuous component).
 
\subsection{Spectrum of the loop/cluster model transfer matrix with no marked loops}
    
\begin{figure}
  \centering
  \includegraphics[scale=.4,angle=-90]{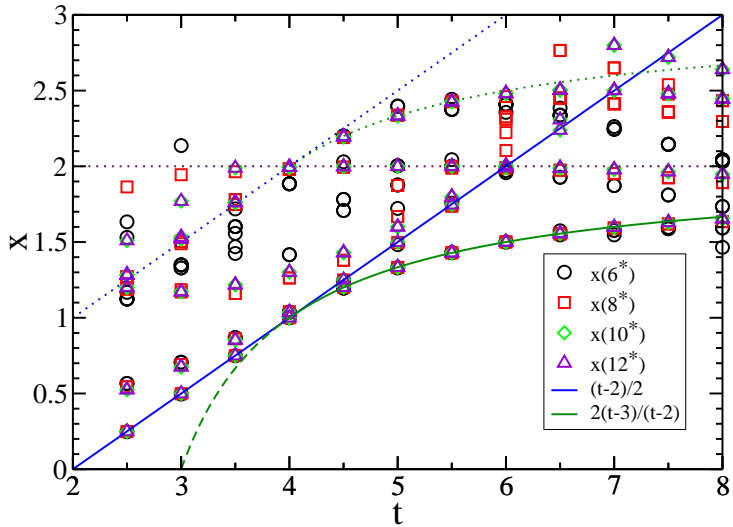}
  \caption{Spectrum of the loop/cluster Potts model transfer matrix in
    the sector with no marked loops/clusters. Two exponents are
    clearly identified (solid lines) as well as descendent levels
    (dotted lines). Axial geometry.}
  \label{num5}
\end{figure}

More generally, the weight $w$ of non-contractible loops is compatible
with charges $e=\pm {1\over t}\pm n$, with $n$ an integer. The choice
$e=e_{0}-1={1\over t}-1$ leads to a conformal weight (defined over the
ground state with $c=2-{6\over t}$)
\begin{equation}
  h={(t-1)^{2}-1\over 4t}={t-2\over 4}
\end{equation}
This is well observed numerically as seen in Fig.~\ref{num5}. We
believe that there is a continuum spectrum on top of this gap, all of
which is truncated when one goes from the Potts transfer matrix to the 
Potts partition function when $t$ is integer.
   
One can also identify numerically a level very well matched by the
formula $h={t-3\over t-2}$. This has no natural explanation within the
XXZ subset, and will be discussed in details below.

We  will argue later that the spectrum is in fact continuous above 
$h={t-2\over 2}$. This is all for $t$ generic of course. For Beraha 
numbers, the continuous component disappears entirely, as well as 
other exponents.

\subsection{Spectrum of the twisted vertex model transfer matrix in the sector 
$S^{z}=0$}

\begin{figure}
  \centering
  \includegraphics[scale=.4,angle=-90]{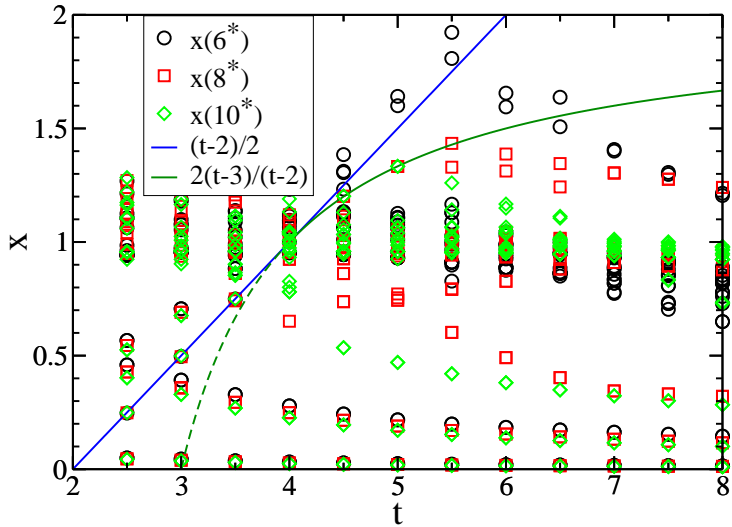}
  \caption{Spectrum of twisted vertex model transfer matrix in the sector $S^{z}=0$. 
    Axial geometry. Note the considerable set of additional 
    levels when compared to Fig.~\ref{num5}. The particular levels
    identified with solid lines in Fig.~\ref{num5} are also present here.
  }
  \label{num6}
\end{figure}

Exponents are shown in Fig.~\ref{num6}. Observe how one sees
considerably more levels than in the loop/cluster model case (see
Fig.~\ref{num5}).
We will argue later that here, the spectrum is in fact continuous
above the ground state $h=0$.

\subsection{Watermelon operators}

\begin{figure}
  \centering
  \includegraphics[scale=.4,angle=-90]{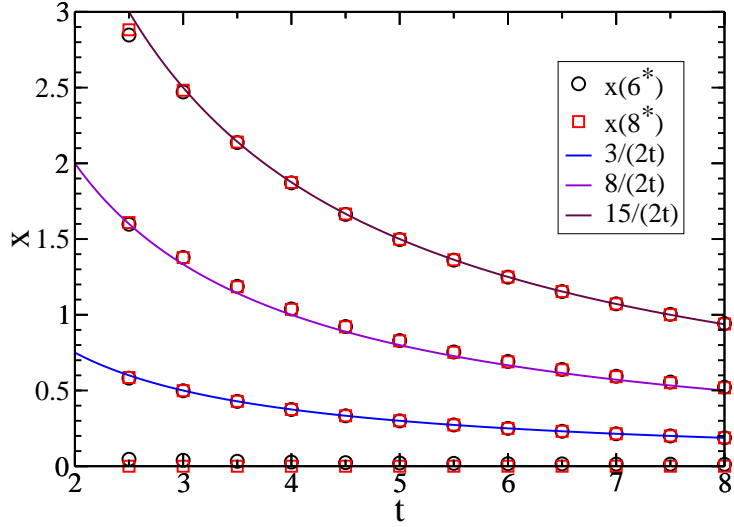}
  \caption{Watermelon exponents on the antiferromagnetic critical
    line, in the loop model with $L=2,4,6,8$ marked lines. Axial geometry.}
  \label{num7}
\end{figure}

These operators describe the properties of $L$ marked lines in the
loop model  (or, for even $L>2$, equivalently by $L/2$ marked
  clusters in the cluster model). As usual, their conformal weight is
obtained by considering the ground state of the sector with
$S^{z}={L\over 2}$ ($L$ even), so, with respect to the ground state of
the theory with central charge $c=2-{6\over t}$, we find
\begin{equation}
  h_{L}={(L/2)^{2}-1\over 4t}
\end{equation}
This is also well checked numerically, as can be seen in Fig.~\ref{num7}. 
  
Notice in particular that the case $L=2$ corresponds to $h_{2}=0$. 
 
\section{RSOS versions at the Beraha numbers}

As commented earlier, the Potts model does present singular behaviour
at the Beraha numbers in the Berker-Kadanoff phase because the largest
eigenvalues of the loop transfer matrix do not contribute to
the partition function. Eigenvalues which actually do contribute are
related, though not identical, to those of another version of the
model, of the restricted solid-on-solid (RSOS) type.  These RSOS
models are particularly interesting to study as they are completely%
\footnote{I.e., they do not even possess a twist.}
local models
of statistical mechanics, and thus should correspond to well defined
conformal field theories, presumably ``minimal'' with respect to some
chiral algebra, and hence easier to identify. 

RSOS models can be defined   whenever $t$ (in 
$q={\rm e}^{i\pi/t}$) is rational, $t=r/s$.  Their variables 
are  
integer heights $h_i=1,2,\ldots,r-1$ living on the union of direct and
dual lattice sites (i.e., again a diagonally oriented square lattice,
but shifted with respect to the medial lattice), subject to the RSOS
constraint $|h_i-h_j|=1$ for neighboring $i$ and $j$.  The Boltzmann
weights are most easily defined by writing the row-to-row transfer
matrix \cite{Pasquier_87,SaleurI}
\be
 T = Q^{N/2}\prod_{i=1}^N \left( \frac{{\rm e}^K-1}{Q^{1/2}}+e_{2i-1} \right)
 \prod_{i=1}^N \left( 1 + \frac{{\rm e}^K-1}{Q^{1/2}} \, e_{2i} \right)
\ee
in terms of Temperley-Lieb generators $e_j$. The latter are represented as
\be
 (e_j)_{h,h'} = \delta(h_{j-1},h_{j+1}) \, \prod_{i\neq j} \delta(h_i,h'_i)
 \, \frac{(S_{h_i} S_{h'_i})^{1/2}}{S_{h_{i-1}}},
\ee
where $S_h = \sin(\pi h s/r)$. 

In the transfer matrix study of the RSOS models, the basis states are
collections of $2N$ heights $h_1,h_1,\ldots,h_{2N}$ belonging alternately
to the direct and dual lattice, and subject to the
constraint $|h_i-h_{i+1}| = 1$. To have
periodic boundary conditions in the $x$-direction we impose $h_{2N+i} =
h_i$.  One can show that ${\rm dim}\ T \simeq \left[2
\cos(\pi/r)\right]^{2N} = Q^N$ when $N \gg 1$.

Of course, the correspondence between the partition function of the 
RSOS model, the partition function of the Potts model, and the 
spectrum of the loop/cluster transfer matrix or twisted vertex model 
transfer matrix is rich and intricate. For instance, the partition 
function of the RSOS model with doubly periodic boundary conditions on 
a torus would read \cite{Pasquier_87}
\begin{equation}
  Z=Q^{S/2} \sum_{E \subseteq \langle ij \rangle}
  Q^{[l(E)-l_0(E)]/2}
  \left( {{\rm e}^K-1\over Q^{1/2}} \right)^{|E|} \
  \sum_{a=1}^{r-1} \left(2\cos{\pi 
      a\over r}\right)^{l_0(E)}
\end{equation}
where the number of non-contractible loops $l_0(E)$ has been singled
out.  This formula indicates that, in the vertex model, $r-1$ sectors
with different values of the twist are now necessary.

In this paper we shall restrict the
study to $t$ integer (i.e., $r=t$ and $s=1$), and use the 
identification of the associated conformal field theory to strenghten 
our general results (on the critical ferromagnetic line for instance, 
the CFT for the RSOS models would be the minimal theory with central 
charge $c=1-6/t(t-1)$).

The critical antiferromagnetic variety does not intersect the selfdual
manifold, and thus corresponds to a {\sl staggered} RSOS model. We are
not aware of any previous study of such model.
   
To proceed, let us define the  four regimes
\begin{eqnarray}
  0&<u<&\gamma~~(i)\nonumber\\
  \gamma&<u<&{\pi\over 2}~~(ii)\nonumber\\
  {\pi\over 2}&<u<&\gamma+{\pi\over 2}~~(iii)\nonumber\\
  \gamma+{\pi\over 2}&<u<&\pi~~(iv)
\end{eqnarray}
The Potts model on the antiferromagnetic critical line involves two
kinds of vertices, with spectral parameters differing by ${\pi\over
  2}$.  Hence, either they lie in regimes $(i),(iii)$ or in regimes
$(ii),(iv)$.

We can go from the Potts model to the RSOS model algebraically by
using a different representation of the Temperley-Lieb algebra (that
is, `quantum group restricting' the vertex model representation).  A
homogeneous spectral parameter $u$ (such as would be obtained on the
selfdual lines) would lead to a homogeneous RSOS model of the type
studied by Andrews, Baxter and Forrester \cite{ABF}. To dispel
confusion, we will reserve the name ABF to homogeneous models, and
denote the restricted version of the vertex model on the
antiferromagnetic line (as well as elsewhere in the phase diagram) as
the RSOS model (in general, staggered).  Where, in the ABF phase
diagram, these models stand is an interesting question.  For $u$ in
regimes $(i)$ or $(iii)$, the ABF model is at the transition point
between its regime $III$ and regime $IV$ (capital letters refer to the
regimes originally defined in \cite{ABF}), while for $u$ in regimes
$(ii),(iv)$, the ABF model stands at the transition point between its
regime $I$ and regime $II$. Based on this observation, we see that we
ought to expect that the antiferromagnetic Potts model, for a given
value of $Q$, can be in two different universality classes, depending
on the anisotropy. The isotropic case is what we are interested in
here: $K_{1}=K_{2}$ corresponds to the SDP (self-dual Potts) models in
regimes $(ii),(iv)$ and thus the ABF models at the transition between
regime $I$ and $II$.  In fact, only the regime $(ii)$ corresponds to a
physical ABF model, and thus was studied in \cite{ABF}.%
\footnote{Note that we also predict from this discussion the existence
  of more critical regimes for the selfdual Potts model itself. Since
  only the isotropic case was usually considered, attention focussed
  on the cases $x_{1}=x_{2}=1$ and $x_{1}=x_{2}=-1$, corresponding
  respectively to $u={\gamma\over 2}$ and $u={\gamma\over 2}+{\pi\over
    2}$, points which lie in regimes $(i)$ and $(iii)$ respectively.
  So the excursions of the selfdual manifold away from the isotropic
  points, in regimes $(ii)$ and $(iv)$ respectively, remains to be
  studied. We suspect that the corresponding universality classes will
  be closely related to the ones of the critical antiferromagnetic
  (hence non-selfdual) Potts model.}
   
Now remarkably, the central charge we have found for the isotropic
Potts model on the antiferromagnetic critical line coincides with the
one of the associated ABF model at the transition between regime $I$
and regime $II$ for $t$ an integer, $t\equiv k+2$. It is therefore
tempting to conjecture that {\sl the RSOS version of the antiferromagnetic
  critical line for $k=t-2$ an integer is in the universality class of
  the ABF models at the transition between regime I and regime II, and
  thus, following \cite{ABF,Huse} is a theory of $Z_{k}$
  parafermions.}  Let us discuss further the evidence for this.
   
\begin{figure}
  \centering
  \includegraphics[scale=.4,angle=-90]{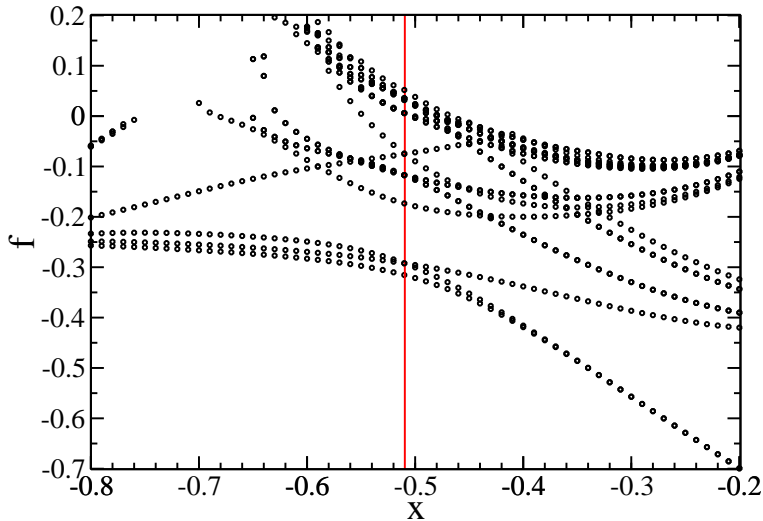}
  \caption{Same as Fig.~\ref{num9} but for the RSOS model.  Remark the
    absence of levels responsible for the BK phase. Some levels are
    identical with those of the cluster model, including the ground
    state on the high-$x$ side of the transition. This ground state
    does not participate any level crossings.}
  \label{num10}
\end{figure}

An immediate objection to this claim might be that the ABF models are
not at a first-order critical point. Indeed, they are not, and the
point is that when one goes from the twisted vertex model to the RSOS
models, the eigenvalues from the untwisted $S^{z}=1$ sector are
discarded ($L=2$ watermelon), as well as the descendent eigenvalues in
the $S^{z}=0$ sector with twist $\alpha={1\over t}$. Therefore, the
second $h=h_{2}=0$ exponent disappears, and there is no reason to
expect a first-order phase transition any longer. This is illustrated 
in Fig.~\ref{num10}. 

Note that in the ABF
language, the extra $Z_{2}$ symmetry which appears on the
antiferromagnetic critical line corresponds to the chiral symmetry of
the $Z_{k}$ models discussed in \cite{Huse}, that exchanges clockwise
and counterclockwise. 
    
\begin{figure}
  \centering
  \includegraphics[scale=.4,angle=-90]{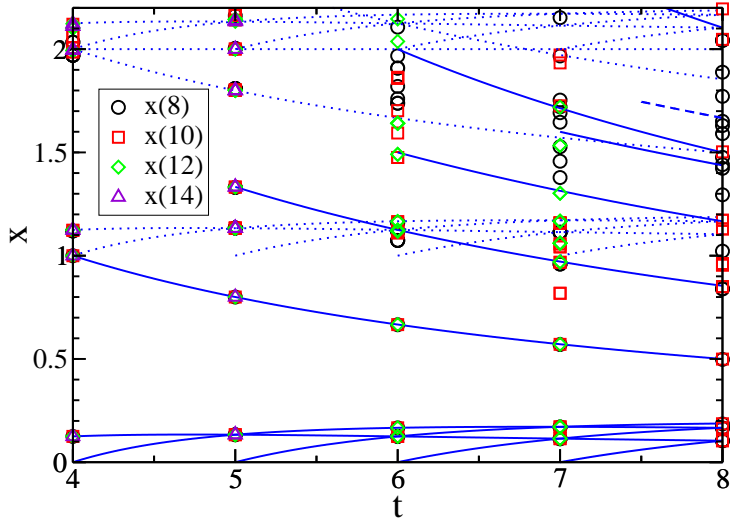}
  \caption{Exponents for the RSOS model on the antiferromagnetic 
    critical line, and comparison with the spectrum of the $Z_{k}$ 
    model. Solid lines show the exponents (\ref{genweight}), while
    dashed lines represent descendents. Diagonal geometry.
  }
  \label{num8}
\end{figure}

The ABF models at the transition between regime $I$ and regime $II$ have
parafermionic exponents, of the general form \cite{Parafermions}
\begin{equation}
  h={l(l+2)\over  4(k+2)}-{m^{2}\over 
    4k}, \quad l=0,1,\ldots,k; \quad -l\leq m\leq l; \quad l-m=0\hbox{ mod }2
  \label{genweight}
\end{equation}
and
\begin{equation}
  h={(k-l)(k-l+2)\over  4(k+2)}-{(k-m)^{2}\over 
    4k}, \quad l=0,1,\ldots,k; \quad l\leq m\leq 2k-l-2; \quad
  l-m=0\hbox{ mod }2\label{othergenweight}
  \label{genweighti}
\end{equation}
 All
these exponents (and only these) are indeed observed numerically in
the RSOS model on the antiferromagnetic critical line: see
Fig.~\ref{num8}.

On the other hand, by the quantum group construction \cite{PS}, the
eigenvalues of the RSOS models are obtained by combining sectors of
the vertex model with vanishing spin and twists $\alpha={(l+1)\over
  (k+2)}$ (recall, $t\equiv k+2$), $l$ integer, $0\leq l\leq k$. The
associated dimension in the XXZ part of the spectrum is obtained by
setting $e={l+1\over k+2}$, and coincides with the one from the sector
with vanishing twist and $S^{z}=l+1$; as such, this eigenvalue
disappears from the loop model in the BK case.  On the
antiferromagnetic line however, this value is still observed: like in
the case $l=0$, there is double degeneracy in the vertex model, so the
loop model still has this exponent.  Observe now that this exponent
(with respect to $c=2-\frac{6}{t}$) agrees with formula
(\ref{genweight}) for the same $l$ and $m=0$.
    
For a given $l$, the parafermionic tower gives weights
(\ref{genweight}).  We recognize there the value coming from the XXZ
subset, from which the quantity ${m^{2}\over 4k}$ has been subtracted.
In particular, in the sector with twist $\alpha= \frac{l+1}{k+2}$, the
lowest lying excitation should not be given by $m=0$ (unless $l=0$)
but by the dimension of the order parameter in the $Z_{k}$ theory,
which corresponds to $l=m$ and reads
\begin{equation}
  h_{l}={l(k-l)\over 2k(k+2)}
\end{equation}
Replacing by the value of $\alpha$ gives the result
\begin{equation}
  h_{\alpha}={((k+2)\alpha-1)(k+1-(k+2)\alpha)\over 2k(k+2)}\label{otherbranch}
\end{equation}
The simplest way to put these results together is that the spectrum
should contain, for any value of $\alpha$, the XXZ value (with respect
to the ground state with $c=2-{6\over t}$)
$h={[(k+2)\alpha]^{2}-1\over 4(k+2)}$. But on top of this, it should
also contain other eigenvalues which have double degeneracy,
(\ref{otherbranch}) being the lowest one for $\alpha={l+1\over k+2}$.
Only for $l=0$, i.e., $\alpha={1\over k+2}$ will the two coincide. This
pattern should presumably extend to $t$ generic.
   
\begin{figure}
  \centering
  \includegraphics[scale=.4,angle=-90]{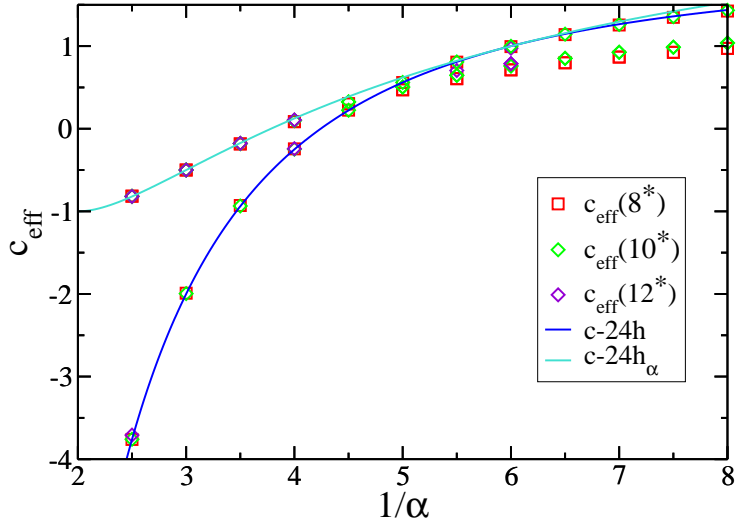}
  \caption{Effective central charge for $t=6$ in the sector with $S^{z}=0$
    and twist $\alpha$. For $\alpha\leq {1\over t}$, the XXZ gap
    determines the effective central charge $c_{\rm
      eff}=c-24h=2-6t\alpha^{2}$ (blue line), while beyond this value,
    the effective central charge is determined by a different gap,
    whose origin will be related to a non-compact boson later, $c_{\rm
      eff}=c-24h_\alpha=-1+{3t\over t-2}(2\alpha-1)^{2}$ (cyan line).
    Axial geometry.}
  \label{num16}
\end{figure}

Numerical study for the leading exponent in the sector of twist
$\alpha$ as a function of $\alpha$ is illustrated in Fig.~\ref{num16}.
One sees that for $|\alpha|<{1\over t}$, the XXZ formula holds, while
beyond this, it has to be replaced by the formula (\ref{otherbranch})
with $t\equiv k+2$, and $t$ now arbitrary, as expected.
   
Exponents with $-l\leq m\leq l$ should also be present, but we have
had difficulties identifying them numerically in general. An exponent
which we have identified however corresponds to (\ref{othergenweight})
with $l=0$ and $m=2$: it is the exponent of the first (most relevant)
parafermionic chiral current, with
\begin{equation}
  h={k-1\over k}
\end{equation}

While we have argued that the spectrum has a continuous component for
the vertex model and the loop/cluster model, there is of course no
such component in $Z_{k}$ models, and therefore all the corresponding
exponents must disappear in the RSOS partition function. This is in
particular the case for the continuous branch in the
sector $l=0$, which starts at $h={t-2\over t}={k\over k+2}$.
   
%
%
%
%
%
%
   
\section{A free field representation for the antiferromagnetic Potts model}

To put all these elements together in the construction of a fully
consistent conformal field theory for the cluster/loop Potts model does not seem straightforward, and
is plagued with the usual difficulties, in particular logarithmic
features---such a fully consistent construction is not available even
in the much simpler case of the ferromagnetic critical line. Like in
the latter case, some progress can be made however by using a free
field representation.
  
This free field representation is inspired by two features. First, we
have deduced the presence of a compact boson and a non-compact boson
from the Bethe ansatz analysis. Second, the value of the central
charge at the Beraha numbers $t=k+2$ coincides (as was observed in
\cite{SaleurI}) with the central charge for $Z_{k}$ models, and for
the latter, a free field representation based on a pair of bosons, one
compact and one non-compact, is well known \cite{Jayaraman}
(see also \cite{GQ,DiQiu}).
   
We thus start, following \cite{Jayaraman} by introducing a pair of
bosonic fields $\phi_{1}$ and $\phi_{2}$, with propagators
\begin{eqnarray}
  \langle\phi_{1}(z)\phi_{1}(w)\rangle&=&-2\ln(z-w)\nonumber\\
  \langle\phi_{2}(z)\phi_{2}(w)\rangle&=&-2\ln(z-w)\nonumber\\
\end{eqnarray}
and a  stress tensor 
\begin{equation}
  T=-{1\over 4}\left(\partial\phi_{1}\right)^{2}+{1\over 
    4}\left(\partial\phi_{2}\right)^{2}+i\alpha_{0}\partial^{2}\phi_{1}
\end{equation}
With a charge at infinity 
$\alpha_{0}= {1\over 2\sqrt{t}}$ for the first boson 
$\phi_{1}$,
the central charge is $c=2-{6\over t}$. 

There are various ways to introduce screening operators in this
theory.  The first choice, which was used in \cite{Jayaraman} uses
both bosons $\phi_{1},\phi_{2}$, and leads to the three currents
\begin{equation}
  J_{1}=\partial\phi_{2}\exp\left[2i\alpha_{0}\phi_{1}\right]
\end{equation}
together with
\begin{equation}
  J_{\pm}=\exp\left[-{i\over 2}\sqrt{t}\phi_{1}\pm{1\over 
      2}\sqrt{t-2}\phi_{2}\right]
\end{equation}
These screening operators, having conformal dimension one, commute with 
the Virasoro algebra. It turns out that they also commute with the 
``parafermionic'' currents
\begin{eqnarray}
  \Psi=-{i\over 2}\left(\sqrt{t\over 
      t-2}\partial\phi_{1}+i\partial\phi_{2}\right)
  \exp\left[{1\over\sqrt{t-2}}\phi_{2}\right]\nonumber\\
  \Psi^{\dagger}=-{i\over 2}\left(\sqrt{t\over 
      t-2}\partial\phi_{1}-i\partial\phi_{2}\right)
  \exp\left[-{1\over\sqrt{t-2}}\phi_{2}\right]
\end{eqnarray}
The vertex operators 
\begin{equation}
  V_{lm}=\exp\left[-i{l\over 2\sqrt{t}}\phi_{1}+{m\over 
      2\sqrt{t-2}}\phi_{2}\right]
\end{equation}
then play a special role: the action of $J_{\pm}$ on $V_{lm}$ is only
well defined for $l$ integer and $l\pm m$ even.  $V_{lm}$ is
annihilated by the corresponding charges $Q_{\pm}$ iff $-l\leq m\leq
l$. Similarly, if we consider the action of powers of $Q_{1}$, it is
well defined only in the case $Q_{1}^{l+1}$ acting on $V_{lm}$, unless
$t$ is rational---we will suppose it is not the case here. Then
$Q_{1}^{l+1}V_{lm}=0$. Finally, one can get outside the range $-l\leq
m\leq l$ by acting with the parafermionic fields, which are also
annihilated by $Q_{\pm}$ and $Q_{1}$.
  
  
%
%
%
%

The case $l=0$ leaves fields obtained by successive action of the
parafermionic fields on the identity, the lowest of which has
dimension $h={t-3\over t-2}$ and is twice degenerate.
  
%
%

We call the fields obtained in this way the first Coulomb gas
contribution. In this sector, $\phi_{2}$ presumably does not
contribute a continuous part, as the corresponding charges are
constrained by the screening operators.
    
Note that we can also consider twisting differently the boson $\phi_{1}$
with $\tilde{\alpha}_{0}={t-1\over 2\sqrt{t}}$. Introducing the usual
screening charges for a one boson theory, $\alpha_{+}\alpha_{-}=-1$,
$\alpha_{+}+\alpha_{-}=2\tilde{\alpha}_{0}$, the corresponding
screening currents read
\begin{equation}
  \tilde{J}_{\pm}=\exp\left[i\alpha_{\pm}\phi_{1}\right]
\end{equation}
The usual Felder construction \cite{Felder} gives the allowed vertex operators
\begin{equation}
  \Phi_{rs}=\exp\left[\left(i \, {1-r\over 2}\, \alpha_{+}+i \, {1-s\over 
        2} \, \alpha_{-}\right)\phi_{1}\right]
\end{equation}
The central charge and the dimensions are not those expected;
$c=2-6{(t-1)^{2}\over t}$ (the additional contribution of one coming
from the boson $\phi_{2}$) and $h={(tr-s)^{2}-(t-1)^{2}\over 4t}$. But
this theory is non-unitary, and admits in particular the negative-dimension
operator with $r=0,s=1$, leading to an effective central
charge $c=2-{6\over t}$ which coincides with the central charge of the
loop/cluster Potts model. The exponents with respect to this effective
central charge are
\begin{equation}
  h={(tr-s)^{2}-1\over 4t}
\end{equation}
and in particular, the choice $r=s=1$ gives the gap $h={t-2\over t}$.
The $L$-leg operators correspond meanwhile to $h_{0,L/2}$ ($L$ even in
the Potts model).
  
In this sector, $\phi_{2}$ is not constrained, and presumably
contributes a continuous spectrum, starting therefore right at the
effective central charge.
  
The ``full Coulomb gas'' is expected to describe the twisted vertex
model: in particular, the degeneracy two of the effective ground state
with $c=2-{6\over t}$ is reproduced in our picture. Some of the levels
of the twisted vertex model were represented earlier.
       
Going to the loop/cluster Potts model involves discarding some levels,
such as the $r=0$, $s=1$ field above. The continuum then should only
start at the true central charge, i.e., the gap $h={t-2\over t}$ with
respect to $c=2-{6\over t}$.
  
When $t$ is rational, going to the RSOS model involves discarding many
more levels, and restricting these Coulomb gases to minimal conformal
field theories.  In the special case $t=k+2$ an integer, the
corresponding minimal model\footnote{We denote by $M_{p,p'}$ the
  minimal model of CFT with central charge $c=1-{6(p-p')^{2}\over
    pp'}$.}  for the second Coulomb gas is $M_{k+2,1}\times$
[non-compact boson $\phi_{2}$] and is actually empty. This leaves
only the first Coulomb gas, whose restriction gives the $Z_{k}$
parafermionic theory. This agrees in particular with our prediction
that the continuous component should disappear at $t=k+2$ an integer.
The case $t$ rational meanwhile should retain some of the continuous
component.
  
\section{The case $N$ odd}

\begin{figure}
  \centering
  \includegraphics[scale=.4,angle=-90]{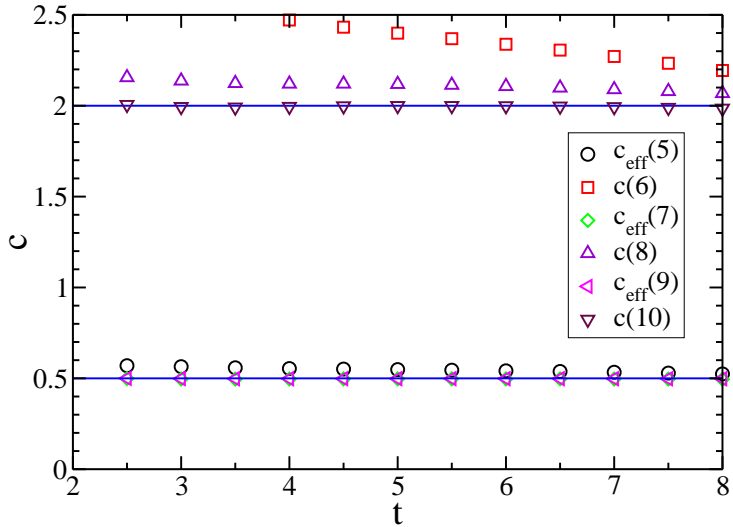}
  \caption{The central charge of the untwisted vertex model on the
    antiferromagnetic critical line for $N$ odd takes the value
    $c_{\rm eff}={1\over 2}$ in contrast with the value $c_{\rm
      eff}=2$ for $N$ even. Diagonal geometry.  }
  \label{num20}
\end{figure}

Numerical study shows that the ground state in the sector with $N$ odd
does not scale like the ground state for $N$ even.%
\footnote{Throughout this section we are referring to the diagonal
geometry only, i.e., the transfer direction is diagonal with respect
to the medial lattice.}
Maybe the best way
to understand this is to start by discussing the untwisted vertex
model, and the associated watermelon operators. We find that the
ground state of the untwisted vertex model exhibits an effective
central charge
\begin{equation}
  c_{\rm eff}={1\over 2}
\end{equation}
independently ot the value of $t$, as illustrated in
Fig.~(\ref{num20}). This is in contrast with the value of $c=2$ we
found earlier for the case $N$ even.
 
The simplest interpretation of this result is that the odd parity acts
as a disorder line, corresponding to a disorder operator with
dimension $\Delta={1\over 16}$. This in turn can be interpreted within
the Coulomb gas by assuming that one of the two bosons sees
antiperiodic boundary conditions in the space direction. The
determination of the watermelon exponents (see below) gives the values
\begin{equation}
  h_{L}={(L/2)^{2}-1\over 4t}+{1\over 16}
\end{equation}
showing that the contribution from the boson $\phi_{1}$ is as in the 
$N$ even case, and thus it must be  the non-compact boson $\phi_{2}$ that 
is twisted, i.e., sees antiperiodic boundary conditions. 
 
Notice that if the field $\phi_{2}$ is twisted in the $N$ odd sector,
the continuous part of the spectrum must disappear, as the twisted
sector of a compact and non-compact bosons are identical. The spectrum
for $N$ odd thus ought to be considerably simpler than for $N$ even, a
fact well in agreement with numerical studies.

\begin{figure}
  \centering
  \includegraphics[scale=.4,angle=-90]{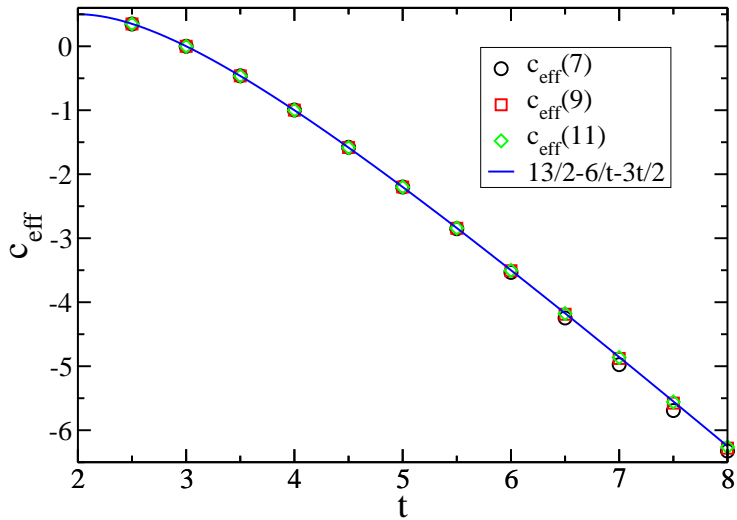}
  \caption{Effective central charge for the loop model on the
    antiferromagnetic critical line in the sector $N$ odd (diagonal
    geometry). The solid line shows the expected result, $c_{\rm
      eff}=c-24
    h_D=\left(2-\frac{6}{t}\right)-24\left(\frac{t-3}{16}\right)$.}
  \label{num17}
\end{figure}

We now turn to the loop/cluster Potts model.
The effective central charge  is found
numerically (see Fig.~\ref{num17}) to be given by a field of dimension
\begin{equation}
  h_{D}={t-3\over 16}\label{observ}
\end{equation}
To interpret the presence of  the weight (\ref{observ}) in general, we observe 
that it can be written 
\begin{equation}
  h_{D_{0}}={1\over 16}+ {t\over 4}\left({1\over 2}-{1\over 
      t}\right)^{2}-{1\over 4t}
\end{equation}
As argued earlier, the ${1\over 16}$ contribution comes from an
antiperiodic sector for the non-compact boson $\phi_{2}$. The
remaining contribution can then be interpreted as the weight for an
electric charge $\tilde{e}_{0}={1\over 2}-{1\over t}$ in the
$\phi_{1}$ theory. Recall that in the $N$ even sector we had charges
$e_{0}={1\over t},1-{1\over t},\ldots$ instead. We do not know the
origin of this additional ${1\over 2}$ contribution to the charge.%
\footnote{Since in the numerical study, the twist for the vertex model
  is exactly $\alpha={1\over t}$, it means that for $N$ odd, this
  twist translates into an electric charge $\tilde{e}_{0}$ for the
  $\phi_{1}$ boson, and a $Z_{2}$ twist for the $\phi_{2}$ boson.
  Moreover, the present value of the charge would correspond, in the
  usual case, to giving to an oriented non-contractible loop, a weight
  $\pm i e^{\pm i\pi/t}$.  This says something about the relation
  between the microscopic arrow degrees of freedom and the bosons
  $\phi_{1}$ and $\phi_{2}$.}



\begin{figure}
  \centering
  \includegraphics[scale=.4,angle=-90]{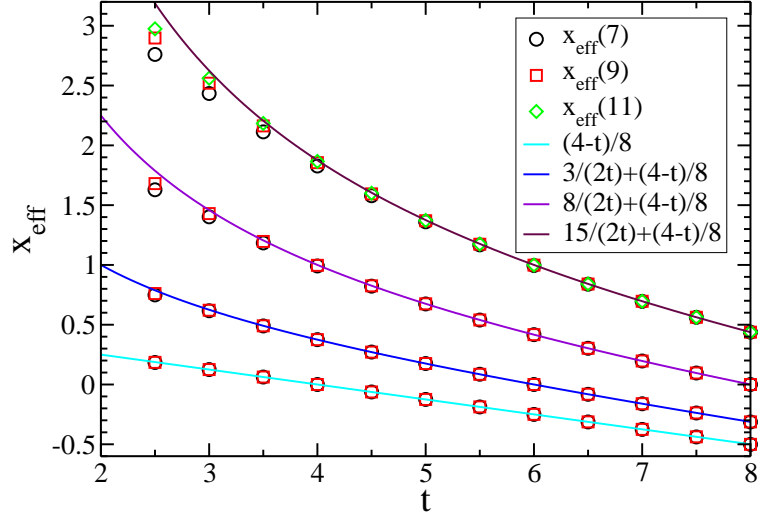}
  \caption{Effective watermelon exponents, in the sector $N$ odd, for
    the loop model along the antiferromagnetic critical line.
    Diagonal geometry.
  }
  \label{num18}
\end{figure}
 
Effective watermelon exponents in the twisted vertex model (and in the
loop model) are then given by measuring the gaps with respect to the
odd-$N$ ground state. This yields
\be
 h_L^{\rm eff} = h_L-h_{D_0} = {(L/2)^{2}-1\over 4t}+{4-t\over 16}
\ee
and is verified numerically in Fig.~\ref{num18}.

For $t=k+2$ an integer, the above results can be fitted in a parafermionic 
picture by introducing the disorder operators (with respect to charge 
conjugation symmetry), whose weights read \cite{ZF_86}
\begin{equation}
  h_{D_{s}}={\left[k-2+(k-2s)^{2}\right]\over 16(k+2)}, \qquad 0\leq s\leq 
  {k\over 2}\label{gendis}
\end{equation}
The largest of these weights is obtained for $s=0$, and coincides 
with (\ref{observ}). 
%
%
%
%
%
On the other hand, it is well-known that correlation functions in the
disorder sectors have to be defined on a two-sheet Riemann sphere,
implying in particular the presence of descendents on half-integer
levels. We expect this feature to extend to $t$ generic, and this
is indeed well verified by the numerics for the first excitations
in the odd-$N$ sector, both for the loop model
and for the twisted vertex model. We spare the reader the 
corresponding figures.



\section{Other exponents of the antiferromagnetic Potts model}
  
We consider first the magnetic exponent.  In the Berker-Kadanoff
phase, it is obtained by putting an electric charge at infinity so as
to cancel the weight of non-contractible loops ($\alpha={1\over 2}$), and
\be
 h_{H}^{\rm BK}={t\over 16}-{(t-1)^{2}\over 4t}.
\label{hHBK}
\ee
The dimension in the untwisted boson theory is thus $\Delta_{H}^{\rm
  BK}={t\over 16}$. The dimension on the antiferromagnetic line,
conjectured first in \cite{SaleurI}, is
obtained by adding a contribution from the non-compact boson, so
\be
 h_{H}^{\rm AF}={t\over 16}-{1\over 4t}-{t-2\over 16}={1\over
  8}-{1\over 4t}.
\label{hHAF}
\ee
This value is easy to understand following the
discussion after formula (\ref{otherbranch}):
the magnetic exponent corresponds
indeed to $\alpha={1\over 2}$ (a value which kills non-contractible 
loops, in relation with the term having $\eta(E)=1$ in (\ref{Zloop})), for which the XXZ value is never valid
as $t\geq 2$ always, while the correct value deduced from the
parafermionic exponents gives $c_{\rm eff}=-1$, in agreement with the
conjectured value of $h_{H}^{\rm AF}$.

\begin{figure}
  \centering
  \includegraphics[scale=.4,angle=-90]{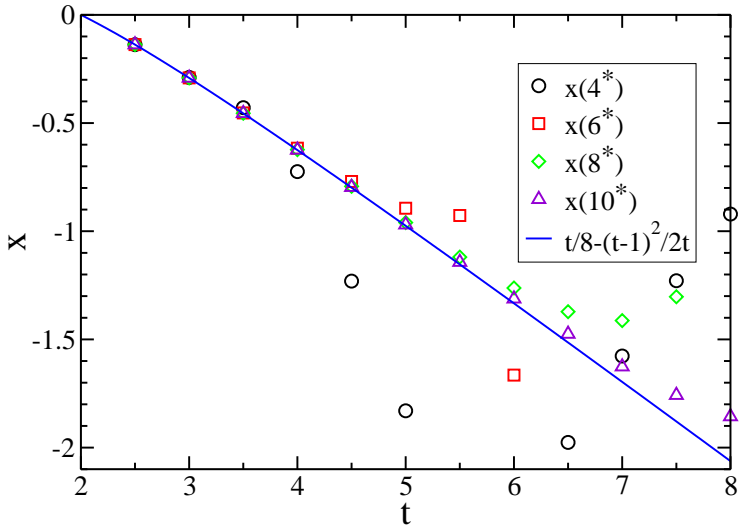}
  \caption{Magnetic exponent of the loop model along the non-physical
    selfdual line. Axial geometry. The solid line represents the
    result (\ref{hHBK}).}
  \label{num23}
\end{figure}

\begin{figure}
  \centering
  \includegraphics[scale=.4,angle=-90]{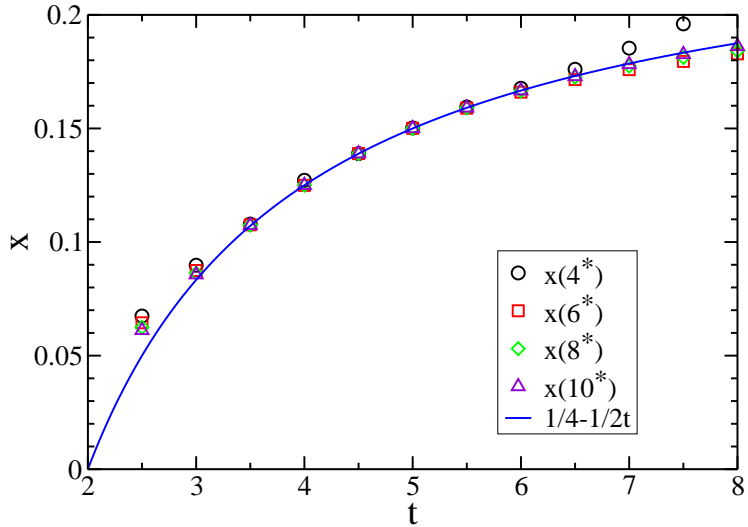}
  \caption{Magnetic exponent of the loop model along the critical
    antiferromagnetic line. Axial geometry. The solid line represents the
    result (\ref{hHAF}).}
  \label{num24}
\end{figure}

The results (\ref{hHBK})--(\ref{hHAF}) for the magnetic exponent
are verified numerically in Figs.~\ref{num23}--\ref{num24}.
  
Next, we consider the thermal exponent. An increase of $K$ moves the
model into a massive phase, while a decrease moves it into the
Berker-Kadanoff (BK) phase. Numerical evidence is that the BK and
massive phases are not related analytically in any way, and that the
largest eigenvalue in the masive phase becomes the smallest (or maybe
one of the smallest, in the sense that it scales with the smallest)
eigenvalue in the BK phase. As a result, the free energy per unit area
for the Potts model at $Q$ generic and the twisted vertex model
exhibits a discontinuity of the first derivative at the
antiferromagnetic critical point, a characteristic of first-order
phase transitions. Nevertheless, we have amply argued that these
models right at the critical point have algebraic decaying
correlations. We are thus in a case of a first-order critical point
with an exponent $\nu=1/2$ all along the line for the twisted vertex
model, and, generically, for the loop/cluster Potts model.
  
\begin{figure}
  \centering
  \includegraphics[scale=.4,angle=-90]{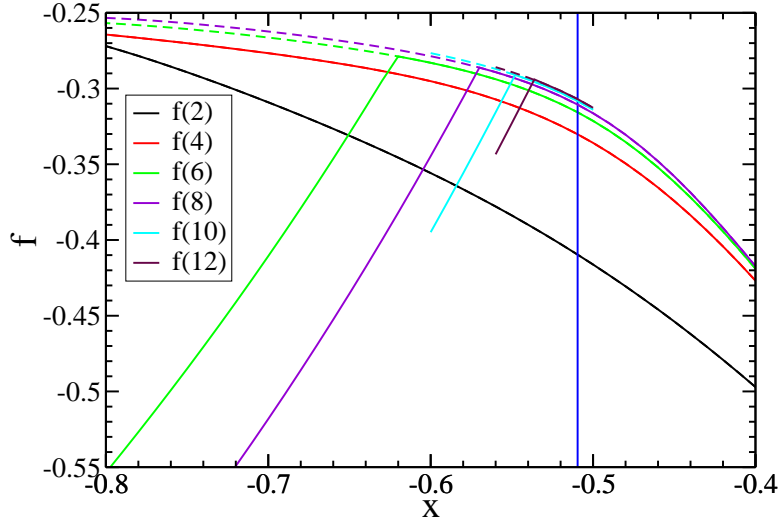}
  \caption{The dashed (resp.\ solid) curves represent the free energy
    of the RSOS model (resp.\ the loop model) at $t=5$ for various
    widths $N$. The two free energies coincide perfectly (and hence,
    also with that of the twisted vertex model) for $x=Q^{-1/2}({\rm
      e}^K-1)$ large enough (whence the dashed curves are `hidden' by
    the solid ones). However, for smaller $x$, the free energy of the
    loop model experiences a sharp singularity with a discontinuity of
    the derivative. As $N$ increases, this singularity moves towards
    the bulk antiferromagnetic transition temperature, here shown as a
    vertical line. Within the BK phase, the free energies of the RSOS
    and the loop models are different, with a jump whose magnitude
    increases as one goes deeper within this phase. Diagonal
    geometry.}
  \label{num14}
\end{figure}

On the other hand, these results do not apply to the loop/cluster
Potts model in the case of $Q$ a Beraha number, that is $t=k+2$
integer.  For such values of $Q$, the free energy does not exhibit a
kink any longer, the eigenvalues crossing the ground state at the
antiferromagnetic critical point being cut-off from the partition
function within the Berker-Kadanoff phase. The same is true for the
RSOS version of the model. These different features are illustrated in
Fig.~\ref{num14}.
   
\begin{figure}
  \centering
  \includegraphics[scale=.4,angle=-90]{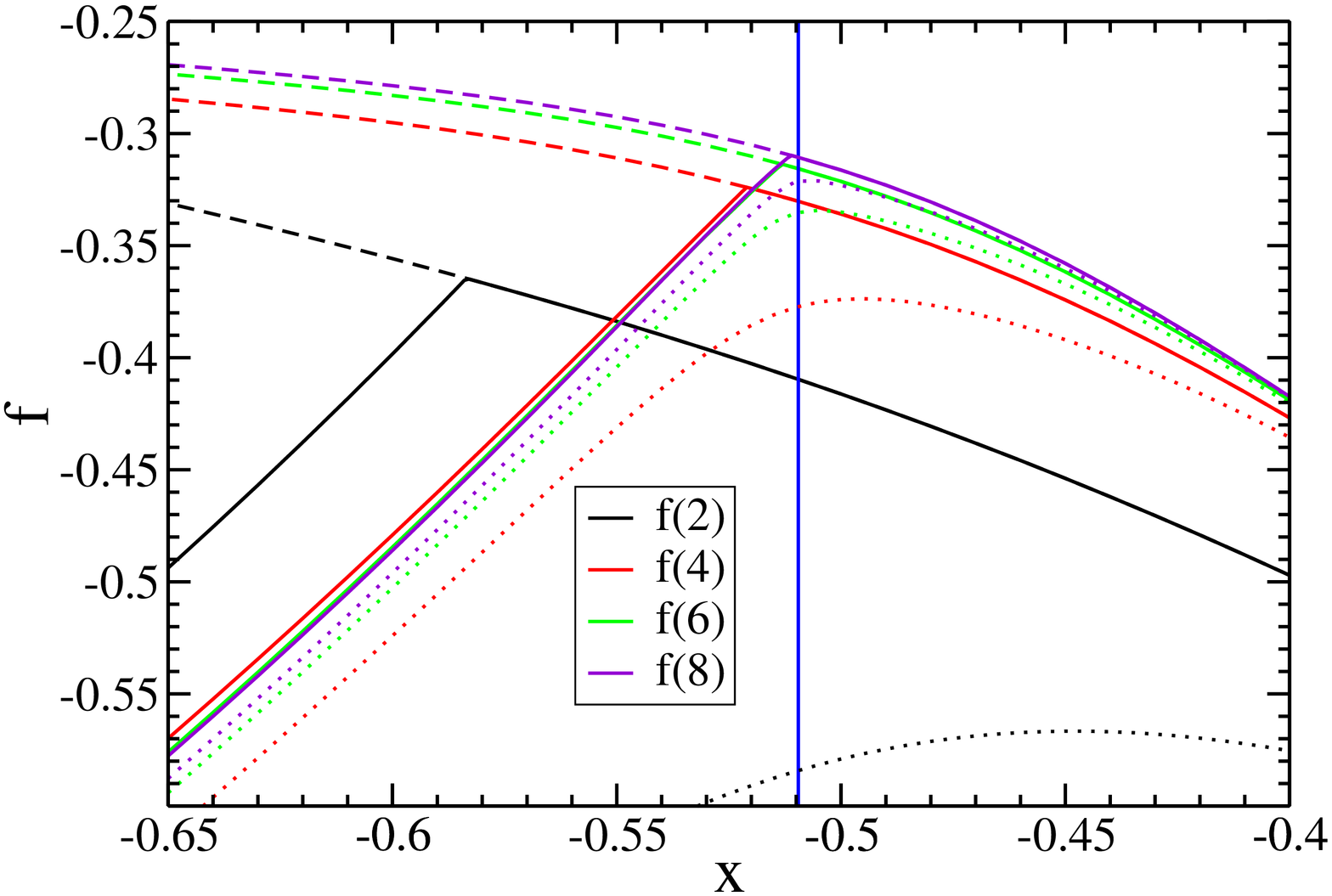}
  \caption{The dotted (resp.\ full) curves represent the free energy
    of the untwisted (resp.\ the twisted) vertex model at $t=5$ for
    various widths $N$. The dashed curves are for the RSOS model, same
    as in Fig.~\ref{num14}. Only the twisted vertex model exhibits
    singular behavior for finite $N$.  However, the free energies of
    the twisted and untwisted vertex model behave similarly in the
    thermodynamic limit, with a first-order critical point on the
    antiferromagnetic critical line. Diagonal geometry.}
  \label{num14a}
\end{figure}

While the existence of the first-order critical point can be pretty 
much argued analytically for the twisted vertex model and the Potts 
model, it is natural to expect a similar transition in the untwisted 
vertex model, since for the vertex model, the free energy per unit 
area should behave ``normally'' and thus be independent of twists. 
This is confirmed by numerical study, as illustrated in Fig.~\ref{num14a}.
	 
At the Beraha numbers, and for the loop/cluster Potts model or the RSOS
model, the thermal operator does not have dimension ${1\over 2}$ any
more.  Rather, the dimension becomes $h={k-1\over k}={t-3\over t-2}$,
the dimension of the parafermion operator.  It is easiest to find out
what happens for the RSOS version of the model, which has the same
largest eigenvalue as the loop/cluster model.  In this case, recall
that the antiferromagnetic critical point is in the universality class
of the $Z_{k}$ theory, and the flow in the Berker-Kadanoff phase
coincides with the flow predicted by Fateev and Zamolodchikov
\cite{FatZam} who studied perturbed parafermionic theories with the
following hamiltonian:
\begin{equation}
  H=H_{k}+\lambda 
  \left({\rm e}^{i\theta/k}\psi\bar{\psi}+
    {\rm e}^{-i\theta/k}\psi^{\dagger}\bar{\psi}^{\dagger}\right)
  \label{FZflow}
\end{equation}
($\lambda$ positive by convention). These authors argued that for $\theta=\pi$ the theory flows, for $k$
odd, to the minimal model with central charge $c=1-{6\over (k+1)(k+2)}$
while it remains massive for $k$ even.

\begin{figure}
  \centering
  \includegraphics[scale=.4,angle=-90]{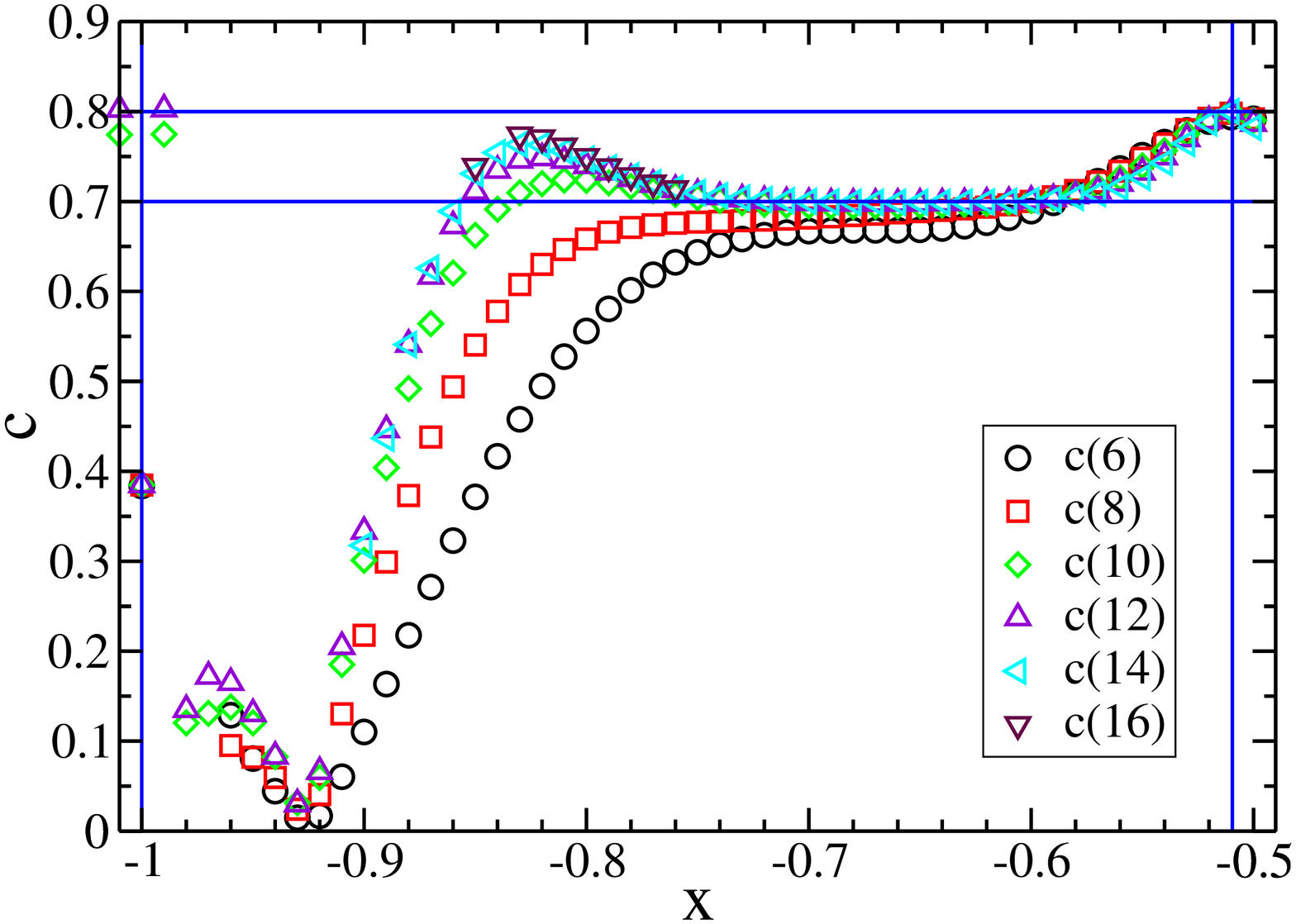}
  \caption{Parafermion to minimal model flow in the RSOS model with
    $t=5$.  The parameter $x = Q^{-1/2}({\rm e}^K - 1)$.  The vertical
    lines show the non-physical selfdual line ($x=-1$) and the
    antiferromagnetic critical point ($x \simeq -0.51$).  The
    horizontal lines are at $c=4/5$ and $c=7/10$, the central charges
    of the relevant parafermion and minimal models. The data indicates
    the existence of a new repulsive fixed point at $x \simeq -0.83$
    in the parafermion model universality class. The minimal model
    fixed point is situated somewhere between this and the
    antiferromagnetic point, and its attractive nature leads to a
    plateau in $c$ in the range $-0.83 < x < -0.51$. The data further
    reveal a $c=0$ fixed point, presumably repulsive, at $x \simeq
    -0.93$. If true, consistency of the RG flow diagram would
    necessitate another attractive fixed point; the data is indeed
    consistent with such a point at $x \simeq -0.87$, again in the
    minimal model universality class.  The interpretation of the data
    in the range $-1 < x < -0.93$ is somewhat uncertain. Right at
    $x=-1$, the data are again well-behaved and converge to $c=-0.385
    \pm 0.001$. We have no explanation for this value at present.
    Diagonal geometry.}
  \label{num11}
\end{figure}

\begin{figure}
  \centering
  \includegraphics[scale=.4,angle=-90]{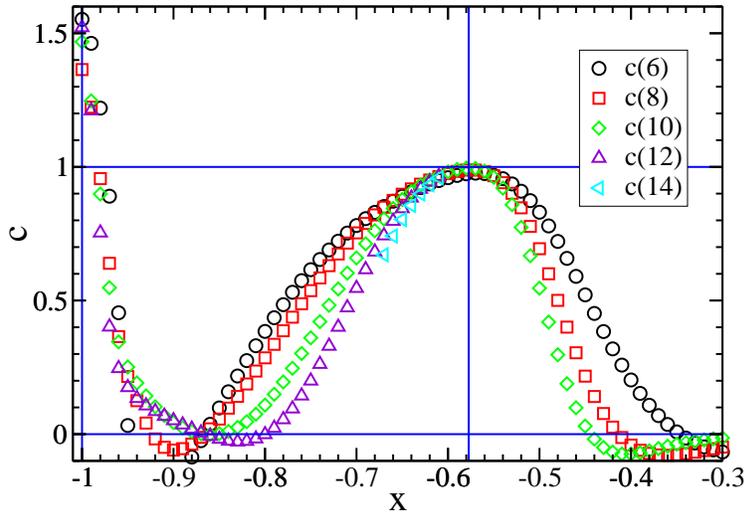}
  \caption{RSOS model phase diagram for $t=6$. The physics is
    different---and simpler---than in the case $t=5$ shown in
    Fig.~\ref{num11}. The $c=1$ parafermion fixed point at the
    antiferromagnetic transition, $x \simeq -0.58$, is still
    repulsive, but the nearest attractive fixed point is now the
    non-critical $c=0$ point at $x \simeq -0.87$. Right at $x=-1$ the
    data converge slowly towards $c=1.6 \pm 0.1$, an unexplained
    value. The model is expected to be massive everywhere, except at
    the two vertical lines. Diagonal geometry.}
  \label{num12}
\end{figure}

\begin{figure}
  \centering
  \includegraphics[scale=.4,angle=-90]{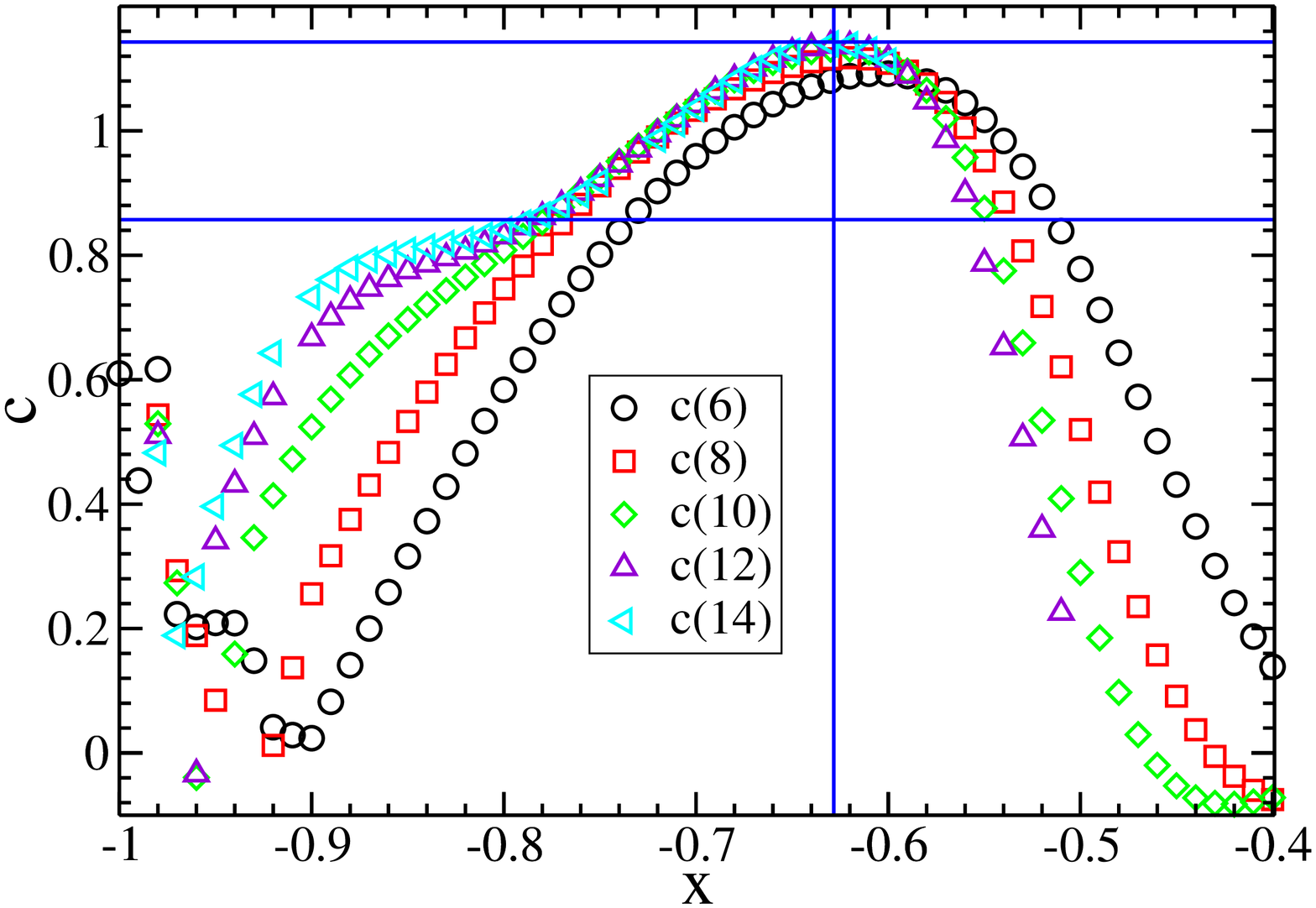}
  \caption{RSOS model phase diagram for $t=7$. The physics is similar
    to the case of $t=5$ shown in Fig.~\ref{num11}. In particular, we
    observe again a parafermion ($c=\frac87$) to minimal model
    ($c=\frac67$) flow. The plateau at $c=\frac67$ develops more
    slowly in $N$, in the form of a `shoulder'. Analyzing the behavior
    at the left edge of that plateau would call for larger system
    sizes than those presented here. Diagonal geometry.}
  \label{num13}
\end{figure}

Numerical evidence for these flows is presented in
Figs.~\ref{num11}--\ref{num13}, where we also give phenomenological
interpretations of other relevant features in the RSOS model phase
diagrams.

\begin{figure}
  \centering
  \includegraphics[scale=.4,angle=-90]{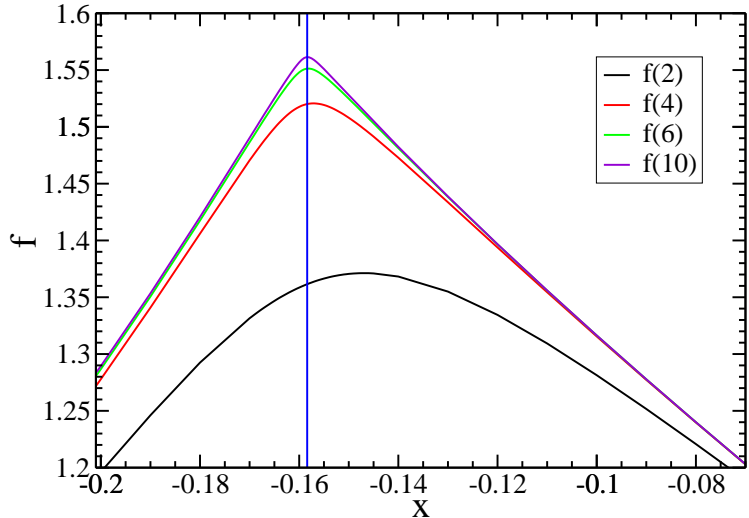}
  \caption{The free energy of the RSOS model for $t={5\over 2}$ also
    develops a first-order singularity in the thermodynamic limit.
    Diagonal geometry.}
  \label{num15}
\end{figure}

Note meanwhile that if $t$ is rational, we can still define an RSOS
model, but its ground state energy should be the same as the one of
the vertex model---that is, exhibit the first-order critical point, in
sharp contrast with what happens for $t$ integer. This is illustrated
in Fig.~\ref{num15} for $t={5\over 2}$.
 
Finally the question remains of how to describe the first-order
critical point within the Coulomb gas.  We have not found the complete
answer to this question, but a probable scenario is that we are
dealing with the perturbation described by Fateev \cite{Fateev},
\begin{equation}
  H=H_{0}+\Lambda 
  \cos(\beta_{1}\Phi_{1})\cosh(\beta_{2}\Phi_{2})\label{fatact}
\end{equation}
where $H_{0}$ is the hamiltonian for a pair of free bosons $\Phi_{1}$ 
and $\Phi_{2}$ as in section 6,  and 
\begin{equation}
  \beta_{1}^{2}-\beta_{2}^{2}=4\pi
\end{equation}
Indeed, consider first the case where the boson $\Phi_{1}$ is twisted
so that the central charge is $c=2-{6\over t}$, and set
${\beta_{1}^{2}\over 8\pi}={t\over 4}$.  By standard Coulomb gas
calculations one finds then the dimension of the perturbation to be
$h={\beta_{1}^{2}-\beta_{2}^{2}\over 8\pi}\pm {1\over 2}=0,1$, i.e., a mix
of a perturbation with dimension zero and one with dimension unity.
Meanwhile, the theory is integrable, and admits non-local conserved
currents, one of them being
\begin{equation}
  J_{\pm}=\partial\phi_{1}\exp\left(\pm{4i\pi\over 
      \beta_{2}}\phi_{2}\right)
\end{equation}
with dimension $h=1-{2\pi\over \beta_{2}^{2}}=1-{1\over t-2}={t-3\over
  t-2}$, the dimension of the parafermionic current. [The other current
meanwhile is
\begin{equation}
  J'_{\pm}=\partial\phi_{2}\exp\left(\pm{4i\pi\over 
      \beta_{2}}\phi_{1}\right)
\end{equation}
with dimensions $h=1$ and $h={t+2\over t}$.] A priori, the integrable 
perturbation involves a mixture of all these operators. What must 
happen is that for $t$ integer, the quantum group truncation leaves 
only the contribution from $J_{\pm}$, leading to the parafermionic 
perturbation. Otherwise, the two relevant operators have dimension 
zero and ${t-3\over t-2}$, presumably in agreement with our 
observation of a first-order critical point for the generic Potts 
model. 

Now in the untwisted theory, the perturbation in (\ref{fatact}) has
dimension $h={1\over 2}$, the one of $J_{\pm}$ is as before, and the
one of $J'_{\pm}$ is $h={t+1\over t}$. One might be concerned by the
absence of a dimension zero operator then, since we have argued that
the untwisted vertex model had a first-order critical point as well.
The point is that the coupling constant $\Lambda$ in (\ref{fatact})
must be identified with the lattice coupling $K-K_{\rm c}$ within the
twisted theory, as multiplying the field of dimension zero in the
twisted theory. Let us call it $\Lambda_{\rm tw}$ for clarity. Then
the physical dimension of $\Lambda_{\rm tw}$ is $[L]^{-2}$. In the
untwisted theory, the perturbation has physical dimension
$[L]^{-2h}=[L]^{-1}$, so the coupling there must also have dimension
$[L]^{-1}$, i.e., be proportional to the square root of $\Lambda_{\rm
  tw}$, or the square root of $K-K_{\rm c}$. A singularity of the free
energy such as $|K-K_{\rm c}|$ turns into $|\Lambda_{\rm tw}|$ i.e.,
$\Lambda_{\rm untw}^{2}$ from dimensional counting, corresponding,
within the untwisted theory, presumably to $\Lambda_{\rm
  untw}|\Lambda_{\rm untw}|$, i.e., a thermal exponent $\alpha=0$, and
thus an exponent $\nu={1\over 2}$, a perturbation of dimension
$h={1\over 2}$ indeed.

Making this more precise is left for future work: it seems clear in
any case that the variable $K$ is not very natural to study the
perturbations of the untwisted theory.

A last observation we can make is that, since the parafermionic 
theories for $t$ integer flow to minimal theories in the BK phase, and 
since the latter are derived from a Coulomb gas with coupling 
$g={t-1\over t}$, it is most likely that excitations for $Q$ generic 
around the ``false ground state''---i.e., the state that crosses from 
the higk $K$ phase, and that would still be the ground state of the Potts 
model for $Q$ a Beraha number---are described by this Coulomb gas, at 
least in the vicinity of the antiferromagnetic transition. Excitations 
around the true ground state meanwhile are described by a Coulomb gas 
with $g={1\over t}$, so both manifolds are in fact present within the 
BK phase!

\section{Conclusion}
  
The least one can say is  that the properties of the antiferromagnetic Potts model
are extremely complex, even for the simple case of the square lattice.
There are definitely loose ends in our study, the most important one
being probably our lack of understanding of the emergence of a
non-compact bosonic degree of freedom. One might understand it at a
very qualitative level by considering the block interactions as in
Fig.~\ref{block}. Summing the arrows on either of the pairs of
incoming or outgoing legs defines dual height variables which are
described by the boson $\phi_{1}$. The states for which the two arrows
have opposite directions do not change the height $\phi_{1}$, and draw
loops on the lattice, which can intersect at vertices. These loops
bear {\sl some} resemblance to the loops of the Goldstone phases in
$O(n)$ models, for which it is known that the continuum limit is
essentially a collection of non-compact bosons and symplectic
fermions. It is thus not so far fetched to expect that the loops in
the antiferromagnetic Potts model are described by a non-compact
boson, which would mean roughly that they are Brownian at large
scales. It would of course be most interesting to elaborate this
picture further.
  
A more straightforward direction of study would be to consider, when
$t$ is rational, the RSOS versions on the antiferromagnetic critical
line. Since when $t=k+2$ is an integer, they are in the universality
class of $Z_{k}$ parafermions, identical with the $SU(2)_{k}/U(1)$
coset model, it is tempting to speculate that they might be related to
$SU(2)$ with a rational level. This will be discussed elsewhere.

The appearance of a line of first-order critical points is intriguing.
Such points are not entirely unheard of. In early works, they were
exhibited for instance in the one-dimensional $q$-state classical
clock model with a topological term \cite{Asorey}. Another case which
is related is the point $H=0$ in the high-temperature loop version of
the low-tempeature Ising model (i.e., the $O(n=1)$ dense loop model
\cite{Nienhuis_82}). Recall indeed that the low-temperature Ising
model is a massive theory as far as the local spin degrees of freedom
are concerned, but nevertheless presents interesting conformal
properties when one reformulates it as a high-temperature loop
expansion. These properties are the same as the ones of the dense
$O(n=1)$ loop model, and are described by a conformal field theory
with vanishing central charge, and, more importantly, a field of
vanishing dimension---the spin operator.  Turning on a magnetic field
leads to a first-order singularity of the free energy, that is, the
model is also at a first-order transition point where the two
manifolds of possible spontaneous magnetization $m=\pm m_{\rm sp}$
intersect. On the other hand, it is not clear how to see the
difference between positive and negative magnetic fields within the
loop formulation, since the field is conjugate to the number of loop
end points, and that number is necessarily even.  Therefore, it is not
clear how to actually observe for instance the first-order singularity
of the free energy.

Interestingly, first-order critical points have also been
independently proposed in a series of papers by Pruisken and
collaborators revisiting the quantum Hall effect transition. These
authors have argued for instance \cite{Pruisken} that in the large-$N$
${\rm CP}^{N}$ model, there are bulk massless degrees of freedom at the
first-order phase transition point. If true, these massless degrees of
freedom could maybe be understood in the $Q$-state Potts model
incarnation as geometrical degrees of freedom, probably non-local in
terms of the original Potts spins.

Finally, the reader might wonder whether the antiferromagnetic
transition discussed here is particular to the square lattice. We
believe on the contrary that it is common to any two-dimensional
lattice. Indeed, it is well established that the properties along the
ferromagnetic transition line are universal, for example by invoking
the standard one-boson Coulomb gas construction. The existence of
another line of transitions for which the thermal operator is
irrelevant (which we have called, for the square lattice, the
non-physical selfdual line) follows by analytic continuation in the
Coulomb gas coupling constant, and leads immediately to the existence
of a Berker-Kadanoff phase, in the region where ${\rm e}^K < 1$. To
isolate this phase from the trivial fixed points at zero and infinite
temperature, we need (at least) on either side a further line of
repulsive fixed points. These lines must necessarily be the loci of
level crossings, and it is thus not far fetched to expect that the
rest of the physics of the first-order critical points (at generic
$Q$) will follow. Moreover, the singularities at the Beraha numbers
follow from the fact that the vertex model transfer matrix can be built in
terms of Temperley-Lieb generators and that this commutes with the
generators of the quantum group, two very universal ingredients.

\begin{figure}
  \centering
  \includegraphics[scale=.4,angle=-90]{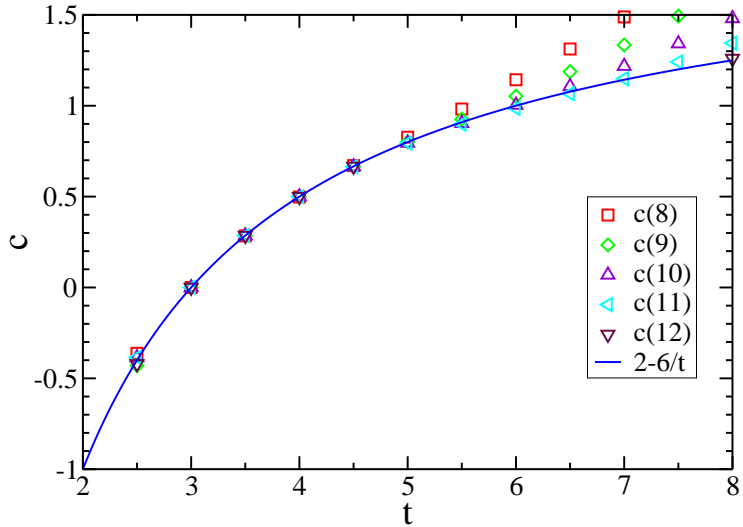}
  \caption{Central charge for the cluster model on the triangular
    lattice.  The transfer direction is perpendicular to a line of $N$
    Potts spins.  All fits are for $N$ which have the same remainder
    modulo 3 (otherwise deviations from the expected result
    $c=2-\frac{6}{t}$ will appear close to $Q=0$ and $Q=4$).}
  \label{num25}
\end{figure}

To test this heuristic argument, let us consider the case of the
triangular lattice. Setting $v={\rm e}^K-1$, the Potts model is
integrable when $v^3+3v^2=Q$ \cite{BTA_78}. When $Q \in [0,4]$ this
describes three branches. The upper one is the ferromagnetic
transition, of course, and by analytic continuation the middle one
should then govern the Berker-Kadanoff phase. The remaining lower
branch is thus a natural candidate for one of the two
antiferromagnetic transition lines. Numerics for the central charge of
the cluster model along the lower branch is shown in Fig.~\ref{num25}
and is indeed well described by $c=2-\frac{6}{t}$. The same results
of course hold by duality for the model on the hexagonal lattice.

\bigskip
\noindent{Acknowledgments}: J.~L.~Jacobsen thanks J.-F.~Richard, J.~Salas,
and A.~Sokal for related collaborations, and SPhT for their warm hospitality.

\begin{appendix}
    
\section{Block spins and the $Q=0$ case}
    
We use for instance the $osp(2/2)$ $\check{R}$ matrix given in
\cite{BassiLeclair}. In the latter reference, a more general
expression is given for the $q$-deformed case. We take the limit
$q\rightarrow 1$. At the same time we scale the spectral parameter
$x=e^{i\gamma v}$, $q={\rm e}^{i\gamma}$, with $\gamma\rightarrow 0$.
We then obtain the expression
\begin{equation}
  \check{R}=P_{1}+{2+v\over 2-v} \, P_{0}-\sqrt{2} {4v\over (v-2)^{2}} \, P_{N}
\end{equation}
The $R$ matrix acts in the tensor product of two fundamental,
four-dimensional representations. The basis states in the latter are
the bosonic states $|1\rangle$, $|4\rangle$ and the fermionic states
$|2\rangle$, $|3\rangle$. The projectors are as follows:

\noindent$(i)$ One-dimensional blocks:
\begin{eqnarray}
  P_{1}=1, \quad P_{0}=0, \quad P_{N}=0, \qquad
  \hbox{ for } |1 \rangle \otimes |1 \rangle
  \hbox{ and } |4 \rangle \otimes |4 \rangle \nonumber \\
  P_{1}=0, \quad P_{0}=1, \quad P_{N}=0, \qquad
  \hbox{ for } |2 \rangle \otimes |2 \rangle
  \hbox{ and } |3 \rangle \otimes |3 \rangle
\end{eqnarray}
\noindent $(ii)$ Two-dimensional blocks:
\begin{equation}
  P_{0}={1\over 2}\left(\begin{array}{cc} 1&- 1\\
      - 1&1\end{array}\right), \quad
  P_{1}={1\over 2}\left(\begin{array}{cc} 
      1& 1\\
       1&1\end{array}\right), \quad
  P_{N}=0
\end{equation}
\noindent for the four pairs of states
\begin{equation}
  \big( |1 \rangle \otimes |2 \rangle, |2 \rangle \otimes |1 \rangle \big), \
  \big( |1 \rangle \otimes |3 \rangle, |3 \rangle \otimes |1 \rangle \big), \
  \big( |2 \rangle \otimes |4 \rangle, |4 \rangle \otimes |2 \rangle \big), \
  \big( |3 \rangle \otimes |4 \rangle, |4 \rangle \otimes |3 \rangle \big)
\end{equation}
and finally
 
\noindent $(iii)$ four-dimensional blocks
 \begin{eqnarray}
   P_{0} &=& {1\over 4}\left(\begin{array}{cccc}
       2&0&0&-2\\
       0&2&2&0\\
       0&2&2&0\\
       -2&0&0&2\end{array}\right)\nonumber\\
   P_{1} &=& {1\over 4}\left(\begin{array}{cccc}
       2&0&0&2\\
       0&2&-2&0\\
       0&-2&2&0\\
       2&0&0&2\end{array}\right)\nonumber\\
   P_{N} &=& {\sqrt{2}\over 4}\left(\begin{array}{cccc}
       1& 1& 1&-1\\
       - 1&-1&-1& 1\\
       - 1&-1&-1& 1\\
       -1&- 1&-1&1\end{array}\right)
\end{eqnarray}
\noindent for the basis states
\begin{equation}
  \big( |1 \rangle \otimes |4 \rangle, |2 \rangle \otimes |3 \rangle,
  |3 \rangle \otimes |2 \rangle, |4 \rangle \otimes |1 \rangle \big).
\end{equation}

\begin{figure}[ht]
  \begin{center}
    \noindent
    \epsfxsize=0.4\textwidth
    \epsfbox{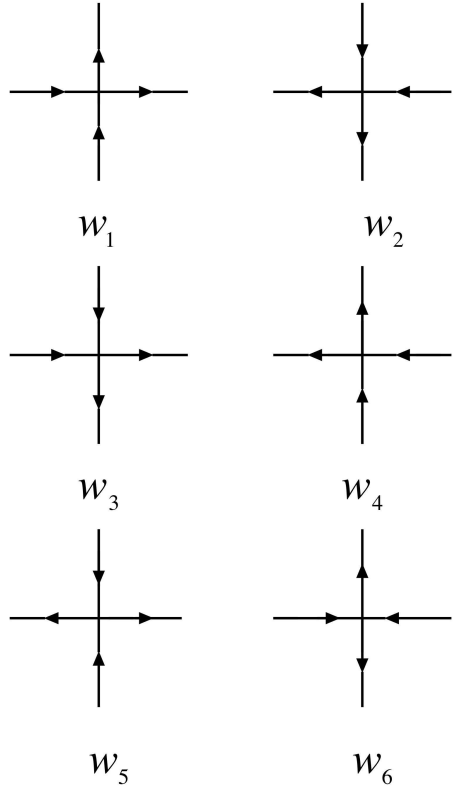}
  \end{center}
  \caption{Weights in the six-vertex model.}
  \label{BW1fig}
\end{figure}

Consider now a different problem where spins $1/2$ interact with the 
well-known six-vertex model $\check{R}$  matrix with anisotropy 
parameter $\gamma$. For the value $u$ of the  spectral parameter 
the Boltzmann weights are represented in Fig.~\ref{BW1fig}, with
\begin{eqnarray}
  w_{1}&=&w_{2}=1\nonumber\\
  w_{3}&=&w_{4}={\sin u\over \sin(\gamma+u)}\nonumber\\
  w_{5}&=&{\lambda\sin\gamma \over \sin(\gamma+u)} \, {\rm e}^{iu}, \quad
  w_{6}={\lambda^{-1}\sin\gamma \over \sin(\gamma+u)} \, {\rm e}^{-iu}
\end{eqnarray}
where $\lambda$ is a parameter at our disposal (a gauge parameter),
and for $\lambda=1$ the $\check{R}$ matrix commutes with
$su(2)_{q}$, with $q=e^{i\gamma}$.

\begin{figure}[ht]
  \begin{center}
    \noindent
    \epsfxsize=0.4\textwidth
    \epsfbox{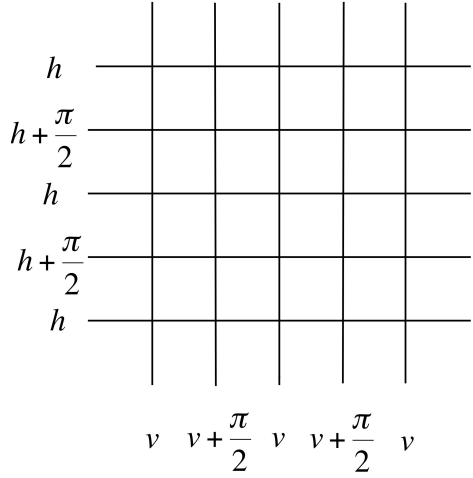}
  \end{center}
  \caption{Staggered vertex model, with alternating parameters
  on horizontal and vertical lines.}
  \label{BW2fig}
\end{figure}

We now consider the following situation. Take a square lattice, and
with every horizontal line associate the parameters $h,h+{\pi\over 2}$
in alternance. With the vertical lines, associate similarly the
parameters $v,v+{\pi\over 2}$ in alternance (see Fig.~\ref{BW2fig}).
The weights at the vertices are given by the foregoing six-vertex
formula with $u$ being the difference of the horizontal parameter and
the vertical parameter for the two intersecting lines. Since the
Boltzmann weights are invariant under a shift of $u$ by $\pi$, we end
up with two kinds of vertices, those for which $u=h-v$ and those for
which $u=h-v\pm {\pi\over 2}$.

We now consider interaction blocks where a pair of horizontal lines
meets a pair of vertical lines. We use these blocks to define a new
interaction matrix, where each link can carry {\sl four} states, made
out of the two spins $1/2$ of the block. In the limit $q\rightarrow i$
($\gamma\rightarrow {\pi\over 2}$), the interaction becomes trivial at
fixed $u$, {\sl but} if we let at the same time $u$ approach zero as
\begin{eqnarray}
  \gamma &=& {\pi\over 2}+\epsilon\nonumber\\
  u &=& \epsilon w
\end{eqnarray}
we claim that we obtain, after inserting appropriate gauge factors, 
the $osp(2/2)$ $\check{R}$ matrix, with the following correspondences
\begin{eqnarray}
  |++ \rangle &\equiv& |1 \rangle\nonumber\\
  |+- \rangle+i\, |-+ \rangle &\equiv& |2 \rangle\nonumber\\
  |+- \rangle-i\, |-+ \rangle &\equiv& |3 \rangle\nonumber\\
  |-- \rangle &\equiv& |4 \rangle
\end{eqnarray}
together with $v=-2w$. 

\begin{figure}
  \centering
  \includegraphics[scale=0.5]{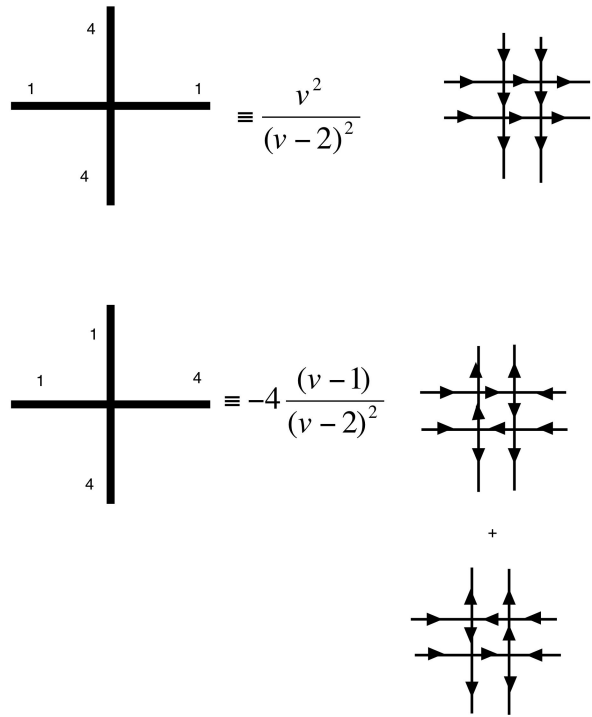}
  \caption{Two examples of the equivalence between $\check{R}$ matrix elements.
  Diagrams on the left are $osp$ vertices together with their 
  Boltzmann weight, and they are obtained from the corresponding 
  ``microscopic'' configurations on the right.}
  \label{BW4fig}
\end{figure}

The check is laborious. We give here a simple example based on the 
fact that
\begin{eqnarray}
  \check{R}_{osp}(|1 \rangle\otimes |4 \rangle)=-{4(v-1)\over (v-2)^{2}} \, 
  |1 \rangle\otimes |4 \rangle+{v^{2}\over (v-2)^{2}} \,
  |4 \rangle\otimes |1 \rangle\nonumber\\
  +{2v\over (v-2)^{2}} \, (|2 \rangle\otimes |3 \rangle +
  |3 \rangle\otimes |2 \rangle)
\end{eqnarray}
This leads among others to the two vertices represented in Fig.~\ref{BW4fig}. 
These vertices are obtained in turn as a sum over configurations of 
the spins $1/2$. For the first, one has the weight 
$$
\left({\sin 
    u\over\sin(\gamma+u)}\right)^{2}\left({\cos 
    u\over\cos(\gamma+u)}\right)^{2}\rightarrow {w^{2}\over (w+1)^{2}} 
$$
and for the second
$$
{\sin^{2}\gamma\over\sin^{2}(\gamma+u)}
{\rm e}^{2iu}-{\sin^{2}\gamma\sin^{2}u\over\sin^{2}(\gamma+u)
  \cos^{2}(\gamma+u)} {\rm e}^{2iu}\rightarrow 1-{w^{2}\over 
  (w+1)^{2}}={1+2w\over (1+w)^{2}}
$$
in agreement with the matrix elements from $\check{R}_{osp}$. 

\begin{figure}
  \centering
  \includegraphics[width=2in]{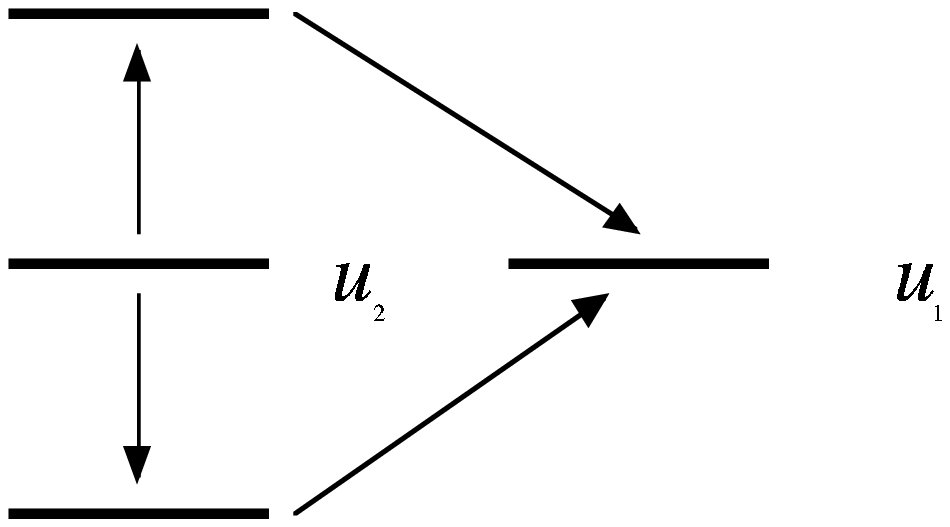}
  \caption{Indecomposable representation $1/2^{\otimes 2}$ for $q=i$.}
  \label{indecrepfig}
\end{figure}

It is interesting to notice that as $q\rightarrow i$, the tensor
product of two fundamental representations of $su(2)_{q}$, which
decomposes as the sum of a one- and a three-dimensional representation
like for $q=1$ when $q$ is generic, becomes indecomposable, and has
the structure shown in Fig.~\ref{indecrepfig}, where
$|u_{1}\rangle=e^{i\pi/4}|+-\rangle+e^{-i\pi/4}|-+\rangle$ and
$|u_{2}\rangle=e^{-i\pi/4}|+-\rangle+e^{i\pi/4}|-+\rangle$, and the
arrows indicate the action of the $su(2)_{q}$ generators:
\begin{eqnarray}
  S^{-}=q^{\sigma^{z}}\otimes \sigma^{-}+\sigma^{-}\otimes 
  q^{-\sigma^{z}}\nonumber\\
  S^{+}=q^{\sigma^{z}}\otimes \sigma^{+}+\sigma^{+}\otimes 
  q^{-\sigma^{z}}
\end{eqnarray}

It turns out that, if one introduces the other generators
\begin{eqnarray}
  \tilde{S}^{-}=q^{-\sigma^{z}}\otimes \sigma^{-}+\sigma^{-}\otimes 
  q^{\sigma^{z}}\nonumber\\
  \tilde{S}^{+}=q^{-\sigma^{z}}\otimes \sigma^{+}+\sigma^{+}\otimes 
  q^{\sigma^{z}}
\end{eqnarray}
at $q=i$, together with the limits $\lim_{q\rightarrow i}{(S^{\pm})^{2}\over 
  q+q^{-1}}$, one obtains the four-dimensional representation of 
$osp(2/2)$. 

\end{appendix}


\begin{thebibliography}{99}
  
\bibitem{PaulMartin} P.P. Martin, ``Potts models and related problems
  in statisticals mechanics'' (World Scientific, Singapore, 1991).
\bibitem{SaleurI} H. Saleur, Nucl. Phys. B {\bf 360}, 219 (1991).
\bibitem{PS} V. Pasquier and H. Saleur, Nucl. Phys. B {\bf 330}, 523 (1990).
\bibitem{EssFSal} F. Essler, H. Frahm and H. Saleur, 
  Nucl. Phys. B {\bf 712}, 513 (2005) [cond-mat/0501197].
\bibitem{JRS} J.L. Jacobsen, N. Read and H. Saleur, Phys. Rev. 
  Lett. {\bf 90}, 090601 (2003) [cond-mat/0205033].
\bibitem{Sokal} J.L. Jacobsen, J. Salas and A.D. Sokal,
 J. Stat. Phys. {\bf 119}, 1153-1281 (2005) [cond-mat/0401026].
\bibitem{Kasteleyn_69} P.W. Kasteleyn and C.M. Fortuin, J. Phys. 
  Soc. Jap. Suppl. {\bf 26}, 11 (1969).
\bibitem{Fortuin_72} C.M. Fortuin and P.W. Kasteleyn, Physica {\bf 57}, 
  536 (1972).
\bibitem{DFSZ} P. Di Francesco, H. Saleur and J.B. Zuber, J. Stat. 
  Phys. {\bf 49}, 57 (1987).
\bibitem{JS_05} J.L. Jacobsen and J. Salas,
  J. Stat. Phys. (in print) [cond-mat/0407444].
\bibitem{BKW_75} R.J. Baxter, S.B. Kelland and F.Y. Wu, J. Phys. 
  A {\bf 9}, 397 (1976).
\bibitem{ReadS} N. Read and H. Saleur, Nucl. Phys. B {\bf 613}, 409 (2001)
  [hep-th/0106124].
\bibitem{BaxterII} R.J. Baxter, Proc. Roy. Soc. London {\bf 383}, 43 (1982).
\bibitem{BaxterI} R.J. Baxter, Studies in Appl. Math. {\bf 50}, 51 (1971).
\bibitem{RS} N.Yu. Reshetikhin and H. Saleur, Nucl. Phys. B
  {\bf 419}, 507 (1994) [hep-th/9309135].
\bibitem{Alcaraz} F.C. Alcaraz, M.N. Barber and M.T. Batchelor, Ann. 
  Phys. (NY) {\bf 182}, 280 (1988).
\bibitem{deVega} H. de Vega and E. Lopes, Nucl. Phys. B {\bf 362}, 261 (1991).
\bibitem{JacSal05} J.L. Jacobsen and H. Saleur,
  Nucl. Phys. B {\bf 716}, 439 (2005) [cond-mat/0502052].
\bibitem{Affleck} I. Affleck, in {\em Fields, strings and critical phenomena},
  Les Houches, 1988, edited by E. Br\'ezin and J. Zinn-Justin
  (Elsevier Science Publishers, Amsterdam, 1989).
\bibitem{Pasquier_87} V. Pasquier, J. Phys. A {\bf 20}, L1229 (1987).
\bibitem{ABF} G.E. Andrews, R.J. Baxter and P.J. Forrester, J. 
  Stat. Phys. {\bf 35}, 193 (1984).
\bibitem{Huse} D. Huse, Phys. Rev. B {\bf 30}, 3908 (1984).
\bibitem{Parafermions} A.B. Zamolodchikov and V.A. Fateev,
  Sov. Phys. JETP {\bf 62}, 215 (1985).
\bibitem{Jayaraman} T. Jayaraman, K.S. Narain and M.H. Sarmadi, Nucl. 
  Phys. B {\bf 343}, 418 (1990).
\bibitem{GQ} D. Gepner and Z. Qiu, Nucl. Phys. B {\bf 285}, 423 (1987).
\bibitem{DiQiu} J. Distler and Z. Qiu, Nucl. Phys. B {\bf 336}, 533 (1990).
\bibitem{Felder} G. Felder, Nucl. Phys. B {\bf 317}, 215 (1989); erratum
  {\em ibid.} {\bf 324}, 548 (1989).
\bibitem{ZF_86} V. Fateev and Al. Zamolodchikov,
  Sov. Phys. JETP {\bf 63}, 913 (1986).
\bibitem{FatZam} V. Fateev and Al. Zamolodchikov,
  Phys. Lett. B {\bf 271}, 91 (1991).
\bibitem{Fateev} V. Fateev, Nucl. Phys. B {\bf 473}, 509 (1996).
\bibitem{Asorey} M. Asorey, J.G. Esteve and J. Salas, Phys. Rev. 
  B {\bf 48}, 3626 (1993) [hep-th/9211048].
\bibitem{Nienhuis_82} B. Nienhuis, Phys. Rev. Lett. {\bf 49}, 1062 (1982).
\bibitem{Pruisken} A.M.M. Pruisken, M.A. Baranov and M. Voropaev, 
  Phys. Rev. Lett. {\bf 505}, 4432 (2003) [cond-mat/0101003].
\bibitem{BTA_78} R.J. Baxter, H.N.V. Temperley and S.E. Ashley, 
  Proc. Roy. Soc. A {\bf 358}, 535 (1978).
\bibitem{BassiLeclair} Z. Bassi and A. Leclair,
  Nucl. Phys. B {\bf 578}, 577 (2000) [hep-th/9911105].


  
\end{thebibliography}
\end{document}